\documentclass[fleqn,usenatbib]{mnras}
\makeatletter
\def\endfigure{\end@float}
\makeatother
\usepackage{newtxtext,newtxmath}
\usepackage{mwe}
\usepackage{caption}
\usepackage{subcaption}

\usepackage[T1]{fontenc}
\usepackage{graphicx}
\DeclareRobustCommand{\VAN}[3]{#2}
\let\VANthebibliography\thebibliography
\def\thebibliography{\DeclareRobustCommand{\VAN}[3]{##3}\VANthebibliography}

\usepackage{graphicx}	
\usepackage{amsmath}	

\usepackage{amssymb}	
\usepackage{threeparttable}
\usepackage{placeins}
\usepackage{graphicx}
\usepackage{wrapfig}
\usepackage{float}
\usepackage{gensymb}
\usepackage{tabularx}
\usepackage{siunitx}
\usepackage{booktabs}
\usepackage[dvipsnames]{xcolor}
\usepackage{tablefootnote}
\usepackage{svg}
\usepackage[normalem]{ulem}
\usepackage[dvipsnames]{xcolor}
\usepackage[normalem]{ulem}
\usepackage[T1]{fontenc}
\usepackage{lastpage}

\DeclareUnicodeCharacter{2212}{-}

\newcommand\fhr{\mbox{$\hspace{0.1cm} \!\!^{\mathrm h}$}}%
\newcommand\fmin{\mbox{$\hspace{0.1cm}\!\!^{\mathrm m}$}}%
\title[Multiwavelength investigations of PKS\,2300$-$18 ]
 {Multiwavelength investigations of PKS\,2300$-$18: S-shaped radio quasar with precessing jets and double-peaked broad emission-line spectrum}

\author[Misra et.al]{
Arpita Misra,$^{1,2}$\thanks{E-mail: amisra@oa.uj.edu.pl}
Marek Jamrozy,$^{1}$
Marek Weżgowiec,$^{3}$
Dorota Kozieł-Wierzbowska$^{1}$
\\
$^{1}$Department of Stellar and Extragalactic Astronomy, Astronomical Observatory, Jagiellonian University, Orla 171, PL-30-244 Krakow, Poland \\
$^{2}$Jagiellonian University, Doctoral School of Exact and Natural Sciences, Krakow, Poland \\
$^{3}$Department of Radioastronomy and Space Physics, Jagiellonian University, Orla 171, PL-30-244 Krakow, Poland  
}

\date{Accepted XXX. Received YYY; in original form ZZZ}

\pubyear{2024}

\begin{document}
\label{firstpage}
\pagerange{\pageref{firstpage}--\pageref{lastpage}}
\maketitle

\begin{abstract}

S-shaped radio galaxy jets are prime sources for investigating the dynamic interplay between the central active galactic nucleus, the jets, and the ambient intergalactic medium. These sources are excellent candidates for studying jet precession, as their S-shaped inversion symmetry strongly indicates underlying precession. We present a multiwavelength analysis of the giant inversion-symmetric S-shaped radio galaxy PKS\,2300$-$18, which spans 0.76 Mpc. The host is a quasar at a redshift of 0.128, displaying disturbed optical morphology due to an ongoing merger with a companion galaxy. We conducted a broadband radio spectral study using multifrequency data ranging from 183 MHz to 6 GHz, incorporating dedicated observations with the uGMRT and JVLA alongside archival radio data. A particle injection model was fitted to the spectra of different regions of the source to perform ageing analysis, which was supplemented with a kinematic jet precession model. The ageing analysis revealed a maximum plasma age of $\sim$ 40 Myr, while the jet precession model indicated a precession period of $\sim$ 12 Myr. ROSAT data revealed an X-ray halo of Mpc size, and from Chandra the AGN X-ray spectrum was modelled using thermal and power-law components. The optical spectrum displaying double-peaked broad emission lines was modelled, indicating complex broad-line region kinematics at the centre with the possibility of a binary SMBH. We present the results of our multiwavelength analysis of the source, spanning scales from a few light-days to a few Mpc, and discuss its potential evolutionary path.

\end{abstract}




\begin{figure*}
    \centering
    \includegraphics[width=1\linewidth]{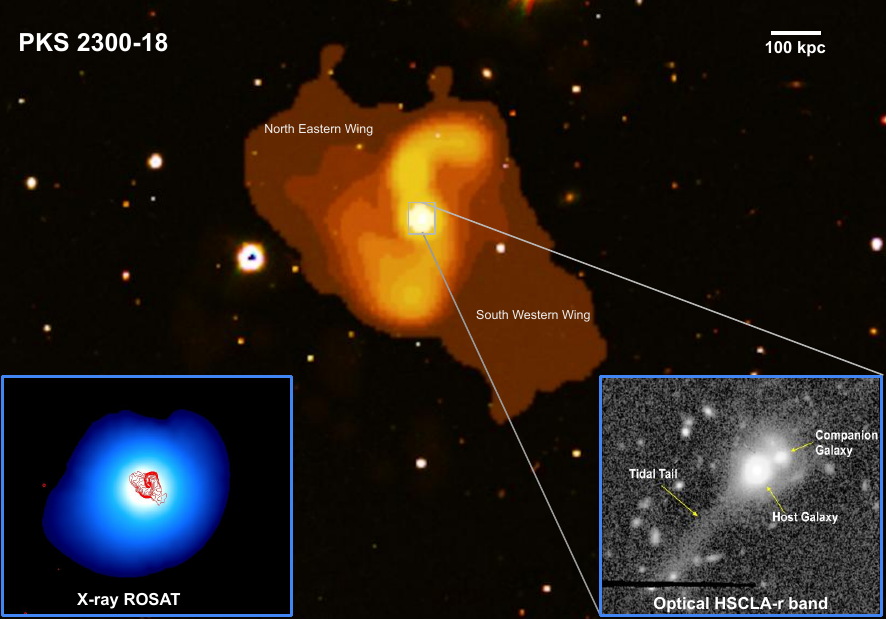}
    \caption{1.4 GHz RACS radio map of PKS\,2300$-$18 (in orange) overlaid on PanSTARRS (\citealt{2016arXiv161205560C}) optical image. The right panel inset is the HSCLA r-band optical image of the host quasar at the centre of the radio source interacting with a companion galaxy and consequentially forming an extended tidal tail. The left inset is the ROSAT map of the quasar showing extended X-ray emission (in blue) with three sigma emission above the noise level, overlaid with contours (in red) from the RACS map at 1.4 GHz (main figure in orange). }
    \label{fig1}
\end{figure*}

\section{Introduction}
\label{sec1}

Supermassive black holes (SMBHs) play a key r\^ole in the formation and evolution of active galactic nuclei (AGNs) at the centres of galaxies. SMBHs accrete surrounding gas and matter, forming an accretion disc where a fraction of the gravitational binding energy is converted into thermal energy and radiation. This process, at times, leads to the production of bipolar relativistic jets that align with the SMBH's spin axis and extend millions of light years before eventually spreading into large-scale radio-emitting lobes (\citealt{1974MNRAS.169..395B}). The jets of these radio galaxies (RGs) penetrate deep into the intergalactic medium, gradually fading in luminosity with increasing distance from the host galaxy or terminating in hotspots at the edges of the diffuse lobes. Standard RGs display a `core-jet-lobe' structure, characterised by symmetric, double-sided radio lobes extending from a central radio core.


RGs are known to exhibit a wide variety of morphological shapes, including twin-tailed jet lobes bent into C-, U-, or V-shaped structures, commonly classified as wide-angle tail and narrow-angle tail galaxies (\citealt{1972Natur.237..269M}; \citealt{1976ApJ...205L...1O}; \citealt{2017AJ....154..169S}; \citealt{2021MNRAS.504.4416D};  \citealt{2023Galax..11...67O}; \citealt{2023JApA...44...17P}). These structures result primarily from ram pressure stripping as the galaxies move through a cluster environment (\citealt{1972ApJ...176....1G}; \citealt{2022MNRAS.516..372B}). Furthermore, some RGs display morphologies resembling X-, S-, or Z-shaped structures, which are classified within the broader class of winged RGs (\citealt{1990Ap&SS.164..131F}; \citealt{2012RAA....12..127G};
\citealt{2019AJ....157..195L}; \citealt{2020MNRAS.495.1271C};  \citealt{2023MNRAS.523.1648M}; {\citealt{2024FrASS..1171101G}}; \citealt{2024MNRAS.530.4902S}). Among these, the S- or Z-shaped radio morphology is comparatively rarer (\citealt{2022ApJS..260....7B}; \citealt{NDUNGU2023101685}), with clear examples of S-shaped morphology observed in sources such as 4C 29.47 (\citealt{1984ApJ...276..472C}), 2MASXJ12032061+1319316 (\citealt{2017MNRAS.465.4772R}), J0644+1043 (\citealt{2024ApJ...969..156S}), and RBS 797 ({\citealt{2024A&A...688A..86U}}).

  The atypical inversion-symmetric `S-shape' in RGs is thought to emerge when standard Fanaroff-Riley type I (FRI) and/or Fanaroff-Riley type II (FRII: \citealt{1974MNRAS.167P..31F}) jets undergo continuous precession and the jets get dragged from their initial position, creating twisted patterns (\citealt{2020MNRAS.499.5765H}). The precession of the jet, driven by a change in the spin-axis direction of an SMBH, gradually tilts the jet axis and ultimately results in an S-shaped morphology over time. A range of possible models for explaining this reorientation or precession have been proposed that include reorientation of the jets due to the presence of another SMBH in the same nucleus (\citealt{1980Natur.287..307B}; \citealt{2001A&A...374..784B}; \citealt{2021ApJ...908..178N}; \citealt{2024MNRAS.530.4902S}) and/or due to a surrounding tilted accretion disk (\citealt{1990A&A...229..424L}; \citealt{2005ApJ...635L..17L}; \citealt{2018MNRAS.474L..81L}; \citealt{2020MNRAS.499..362C}). Galaxy mergers, which are a common evolution trajectory for most galaxies following hierarchical structure formation (e.g. \citealt{1977ApJ...217L.125O}; \citealt{1993MNRAS.262..627L}), have an important r\^ole behind forming such radio morphologies. Mergers drive the processes that naturally lead to the formation of SMBH binaries and complex gas dynamics at the centre of galaxies, which contribute to jet precession. 
  
 One of the most compelling examples linking S-shaped radio jets to precession is the quasar PKS\,2300$-$18 (other names PKS\, B2300$-$189, QSO 2300-189 and J2303$-$1841). This source is notable for its striking inversion symmetry in radio and is considered one of the clearest examples of precessing jets observed in quasars (\citealt{1984MNRAS.207...55H}). PKS\,2300$-$18 has S-shaped jets surrounded by diffuse low surface brightness wings that contain old plasma. In RGs, such emission can be detected as the phase of nuclear activity in a radio AGN can range from several million to several billion years. Over time, particles in the radio lobes lose energy through synchrotron radiation and inverse-Compton scattering, leading to the gradual fading of the lobes and jets. However, these radio sources still remain traceable long after their emitting electrons were last accelerated to relativistic energies, thus preserving a comprehensive historical record of jet activity.

\begin{table*}
\caption{Details of uGMRT and the JVLA dedicated observations of PKS\,2300$-$18 analysed in this work.}
\label{tab:landscape1}
\begin{tabular}{ccccccccc}
\hline
Telescope & Frequency range & Central Freq. &  Proposal code & Date of observation & TOS &  Resolution & PA & rms \\

&  (MHz)  & (MHz) &  &  & (hrs) & (arcsec$\times$arcsec) & (\degree)   & (mJy \( \text{beam}^{-1} \))  \\
(1) & (2) & (3) & (4) & (5) & (6) & (7) & (8) & (9)  \\
\hline
\\
uGMRT Band--2   & 120–250 & 183 &  40\_016    & 2021 Aug 14   & 4.1 & 18.6$\times$11.2 & 36.8 & 1.28\\
\\
uGMRT Band--3  & 250–500 & 320 &  40\_016    & 2021 Aug 17   & 4.0 & 12.2$\times$7.2 & 23.7 & 0.10\\
\\
uGMRT  Band--3   & 250–500 & 400 &   40\_016  & 2021 Aug 17  & 4.0 & 8.9$\times$5.6 & 32.8 & 0.09\\
\\
uGMRT Band--4    & 550–850 &607 &  39\_018  & 2020 Dec 11   & 4.0 & 6.0$\times$3.7  & 27.3 & 0.11\\
\\
JVLA C band D--conf & 4000--8000 & 6000 &  22A\_091  & 2022 Aug 31  & 0.2 & 14.4$\times$9.0 & 1.8 & 0.03 \\
\\

\hline
\end{tabular}
\label{table1}
\end{table*}

In this article, we examine the radio emission of PKS\,2300$-$18, which features low surface brightness radio wings that span $\sim$ 5 arcmin and S-shaped jets that extend $\sim$ 3 arcmin in the plane of the sky. Identified by \citet{1966AuJPh..19..559B}, the source displays a concave radio spectrum, i.e. steep at low frequencies and flat at high frequencies (\citealt{1984MNRAS.207...55H}). This spectrum is typically observed in sources with a compact, flat-spectrum core surrounded by steep spectrum extended emission.  PKS\,2300$-$18  is located at  RA: $23\fhr03\fmin02\fs97$ and Dec: $-$18\degr41\arcmin25.\arcsec8 (J2000.0) with a redshift of z = 0.1283 (\citealt{1978MNRAS.185..149H}). The quasar, with an absolute B-band magnitude of -20.51, is identified by its broad emission lines. 
Optical i-band images from the Hyper Suprime-Cam Legacy Archive (HSCLA; \citealt{2021PASJ...73..735T}) reveal a companion galaxy located 14 kpc away, with both galaxies embedded in a common envelope of gas and dust (\citealt{1984MNRAS.207...55H}). The quasar and its companion are undergoing a merger resulting in a long tidal tail extending from the host and a small c-shaped tidal arm wrapped around the companion galaxy, as seen in Fig.~\ref{fig1} (bottom right insert). This source is not associated with any known galaxy group or cluster. The quasar shows optical variability, as observed in the All-Sky Automated Survey for Supernovae  (ASAS: \citealt{2020MNRAS.491...13J}) from 2013-2018 and in Zwicky Transient Facility monitoring (\citealt{2019PASP..131a8002B}). Furthermore, the radio core of PKS\,2300$-$18 also shows strong variability between 5-40 GHz, seen in The Planck Australian Telescope Compact array (Planck-ATCA) Co-eval Observations (PACO: \citealt{2013MNRAS.428.1845B}). A more detailed discussion of this is presented in Section~\ref{sec3.3}.
The host is also identified as a gamma-ray-emitting RG, as detected by the Fermi satellite in 4FGL-DR2 (\citealt{2020arXiv200511208B}; \citealt{2023ApJS..265...60C}). A morphologically similar source to PKS\,2300$-$18, J1257+1228 was discussed in \citet{2020ApJS..247...53K} paper, which has an X-shaped radio morphology.

In this study, we performed a multiwavelength analysis of PKS\,2300$-$18, examining its radio emission, optical spectrum, and X-ray emission. We used archival and dedicated radio observations covering frequencies from 183 MHz to 40 GHz to investigate the galaxy in detail. By analysing its radio spectra, we assessed the ageing of the electron population in various regions of the source and gave an account of its structural evolution. In addition, we modelled the optical and X-ray spectra of the AGN. In this work, Section~\ref{sec2} provides a comprehensive overview of the radio observations along with details of the optical and X-ray observations. Sections~\ref{sec3},~\ref{sec4}, and~\ref{sec5} present the results of our radio, optical, and X-ray analyses, respectively. A detailed discussion is presented in Section~\ref{sec6} and the conclusions are drawn in Section~\ref{sec7}.

All absolute quantities in this work were calculated for a $\Lambda$CDM universe with $H_0$ = 70 km s${^{-1}}$Mpc${^{-1}}$, $\Omega_{\rm{m}}$ = 0.3, and $\Omega_{\rm{\Lambda}}$ = 0.7. Using the host galaxy redshift, the conversion scale translates to 2.291\,kpc/\arcsec,  used in \citet{2006PASP..118.1711W}.

\begin{figure*}
    \centering
    \includegraphics[width=1\linewidth]{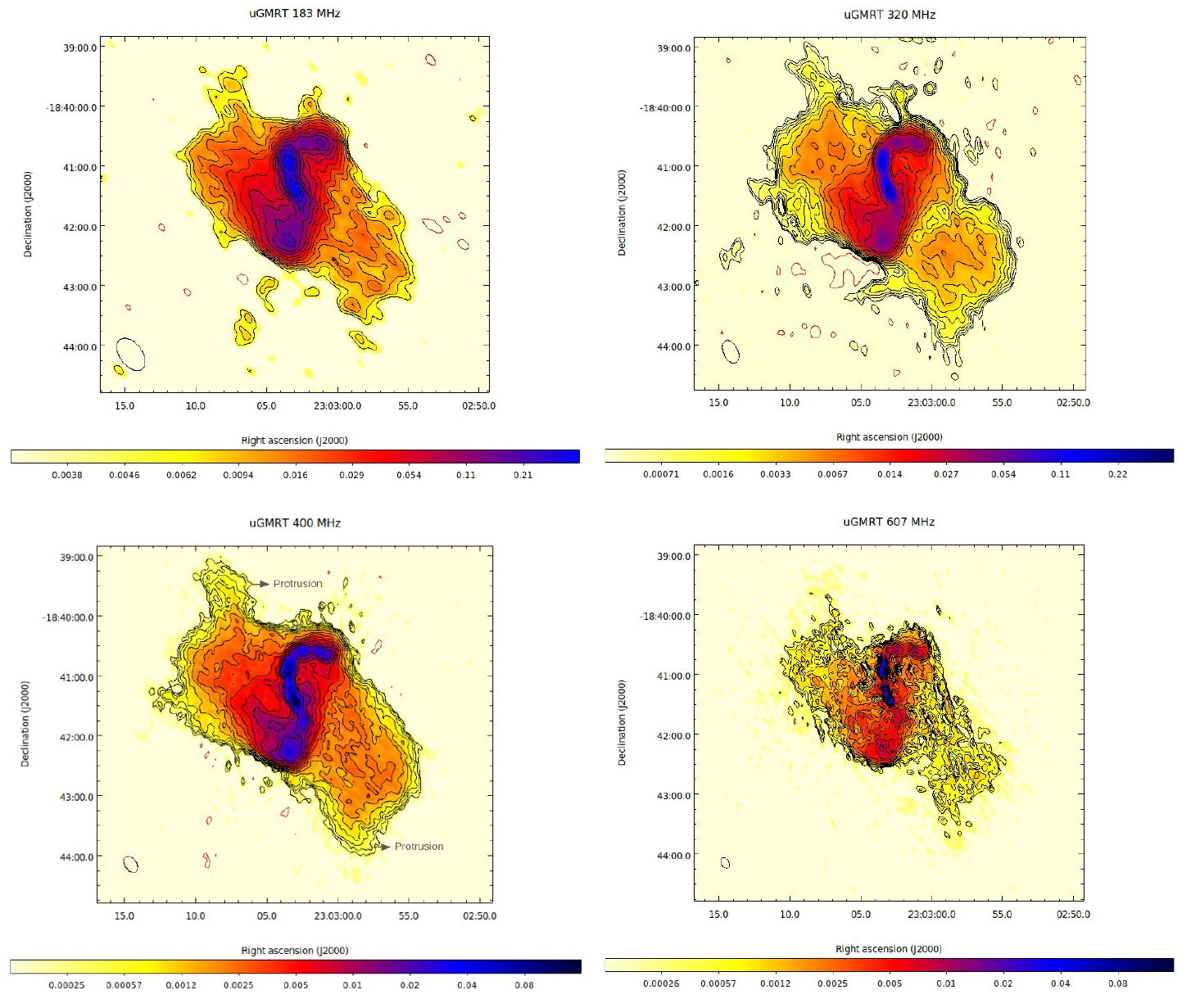}
    \caption{Low-frequency maps of PKS\,2300$-$18 obtained from uGMRT observations at 183, 320, 400 and 607 MHz. The contour levels are spaced by a factor of $\sqrt{2}$ and the first contour is at 3 $\times$ rms level. The first negative 3 × rms level is marked with red contours. The rms values and beam sizes of all the above maps are given in Table~\ref{table1}. The relative sizes of the beam are indicated by the ellipse at the bottom left corner of each image. The colour gradient represents flux density values in Jy \( \text{beam}^{-1} \).}
    \label{fig2}
\end{figure*}

\begin{figure*}
    \centering
    \includegraphics[width=1\linewidth]{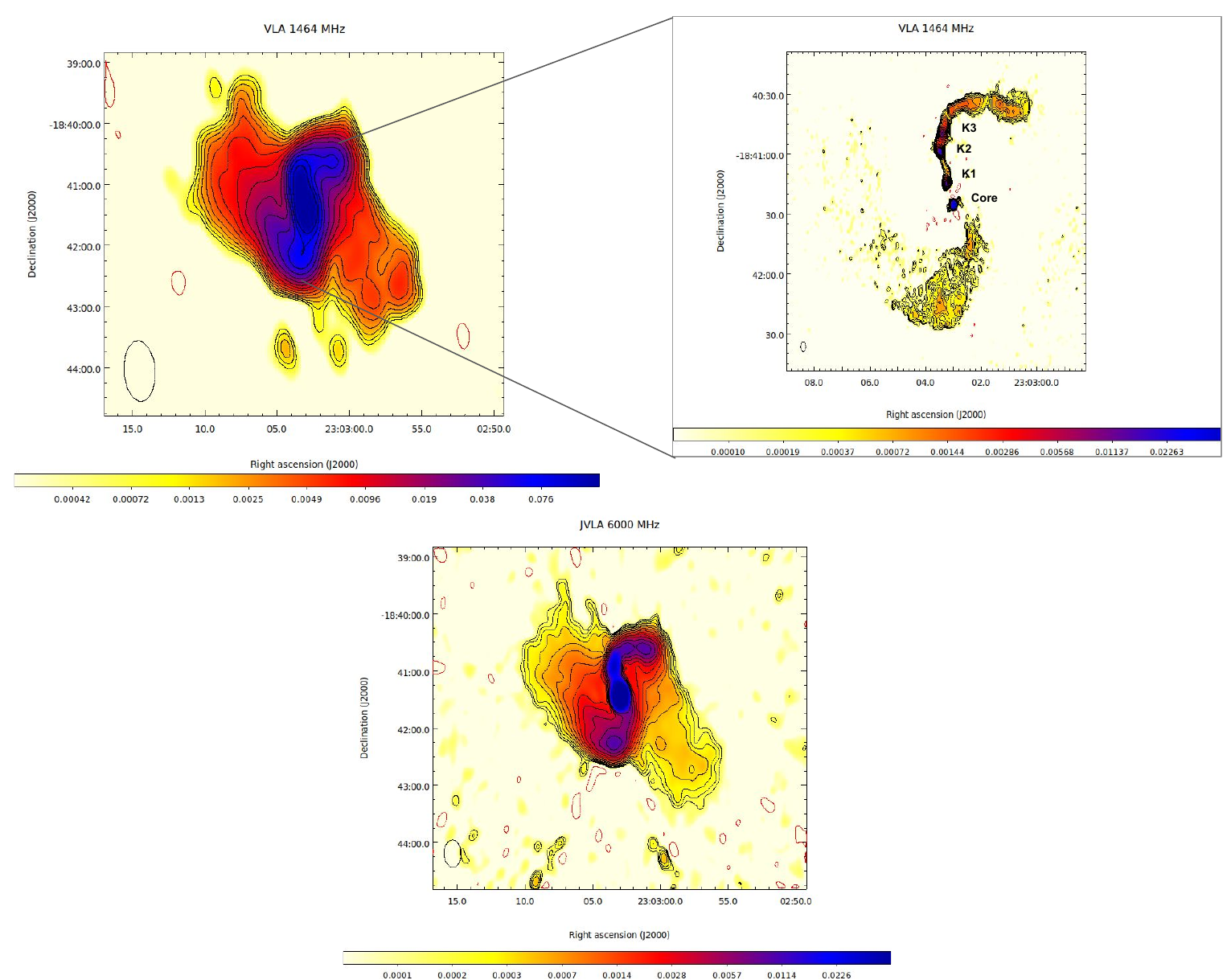}
    \caption{High-frequency maps of PKS\,2300$-$18 obtained from VLA observations at 1.4 GHz (archival) and 6 GHz (dedicated). The contour levels are spaced by a factor of $\sqrt{2}$ and the first contour is at 3 $\times$ rms level. The first negative 3 × rms level is marked with red contours. The beam size of the archival VLA 1464 MHz (C config) map is 29.7$\times$15.2 arcsec and rms value is 0.27 mJy \( \text{beam}^{-1} \) (rms values and beam sizes of 6000 MHz map and 1464 MHz A and C configuration maps are given in Tables~\ref{table1} and ~\ref{tabnew}). The relative sizes of the beam are indicated by the ellipse at the bottom left corner of each image. The colour gradient represents the flux density values in Jy \( \text{beam}^{-1} \).}
    \label{fig3}
\end{figure*}

\section{Observations and data reduction}
\label{sec2}


\subsection{Radio observations}
\label{sec2.1}
 
 To elucidate the radio morphology of PKS\,2300$-$18, high-resolution multifrequency radio maps were obtained using dedicated observations from the upgraded Giant Metrewave Radio Telescope (uGMRT), the Karl G. Jansky Very Large Array (JVLA), and publicly available survey data.  In the following subsections, we provide a detailed overview of the observation methodologies, the surveys utilised, and the data processing techniques implemented to generate the uGMRT and JVLA maps presented in Fig.~\ref{fig2} and Fig.~\ref{fig3}.

\subsubsection{uGMRT observations}
\label{sec2.1.1} 

PKS\,2300$-$18 was observed using the uGMRT wideband receiver (GWB) with 2048 channels in three frequency bands, band 2 (125$-$250 MHz), band 3 (250$-$500 MHz), and band 4 (550$-$850 MHz), concurrently with the narrowband receiver (GSB), details are given in Table.~\ref{table1}. For the observations, the usual scheme of looping the phase calibrator with the target source was adopted, with the flux-density calibrators observed for $\sim$10 mins at the beginning and the end of observations. Observations of the target source from bands 2 -- 4 were made using the flux density calibrator 3C\,48. In band 4, J2246$-$121 as the phase calibrator was observed for 5 mins alternating with the target source in the loop. At bands 2 and 3, J2225$-$049  was observed as the phase calibrator for 5 mins between cycles.\\

 The reduction of GSB narrowband data for all observed frequencies was carried out using the Source Peeling and Atmospheric Modeling pipeline (SPAM; 
 \citealt{2009A&A...501.1185I}; \citealt{2014ascl.soft08006I}). SPAM is a Python-based extension to AIPS (\citealt{2003ASSL..285..109G}) that reduces high-resolution, low-frequency radio interferometric observations. It follows direction-dependent calibration, modelling, and corrections for the dispersive phase delay that are mainly of ionospheric origin.  The data were corrected for strong radio frequency interference (RFI) followed by standard flux density, and bandpass calibration. Subsequent cleaning was performed using the Cotton-Schwab algorithm (\citealt{1984AJ.....89.1076S}; \citealt{1999ASPC..180..357C}). This algorithm is a modified version of CLEAN deconvolution (\citealt{1974A&AS...15..417H}; \citealt{1980A&A....89..377C}) and enables the deconvolution of multiple facets simultaneously where each facet is deconvolved using a different dirty beam. The wideband GWB data was also reduced using SPAM, where the data was first split into 6-9 subbands. For the calibration of the individual subbands, a skymodel was extracted from the initial narrowband image using the Python Blob Detector and Source Finder (PyBDSF;  \citealt{2015ascl.soft02007M}), which is a Python-based algorithm that can decompose an image into a set of Gaussians, shapelets, or wavelets. After calibration, the final imaging was performed using the w-stacking algorithm WSclean (\citealt{offringa-wsclean-2014}), which is an alternative to the w-projection algorithm in which the uv samples are not convolved with a w-term correcting kernel before the FFT, but are instead corrected with a multiplication after the FFT.  In WSclean, a multi-scale, multi-frequency wideband deconvolution approach (\citealt{2017MNRAS.471..301O}) was used to image each band. The multi-frequency deconvolution feature for wideband enables the simultaneous cleaning of channels by considering spectral variation and using each channel’s specific PSF for deconvolution.
Here the \citet{1995AAS...18711202B} weighting scheme was used to suppress the sidelobes in the point spread function (PSF). A robustness parameter of 0 was used, as an ideal between natural and uniform scales to preserve both large-scale structures and high-resolution details.  The resultant images were primary beam corrected using the AIPS task PBCOR. The final uGMRT maps are presented in Fig.~\ref{fig2}.

\subsubsection{VLA observations}
\label{sec2.1.2}

Dedicated observations of PKS\,2300$-$18 were performed in C band using D-configuration of the JVLA, with bandwidth of 4096 MHz and 2048 channels. A single scheduling block was used, consisting of calibrator (both flux and phase) and source scans. The data were processed with the Common Astronomy Software Applications (CASA; \citealt{2022PASP..134k4501C}) package version 6.5.6.22. All data affected by RFI were flagged manually. The radio sources 3C\,286 and J2246$-$1206 were used for the flux density scale and phase calibrations, respectively. The deconvolution of the interferometer point spread function (PSF) from the dirty map was performed with the use of the CASA task {\sc tclean}. Within this task the Clark CLEAN algorithm was used and to ensure the best dynamic range, that is, the sensitivity to faint large-scale emission, multiscale clean with Briggs weighting was used (robust parameter set to 0). Due to a significant brightness of the source core region, as well as its complex structure, severe interference from the sidelobes caused the need for self-calibration (a total of three runs), which was performed in phases only.

To obtain high angular resolution maps of PKS\,2300$-$18, necessary for the analysis presented in Section~\ref{sec3},  L- and C-band data of the target from the VLA archive were retrieved and reduced using AIPS (Bridle \& Greisen 1994, NRAO AIPS Memo 87 \href{http://www.aips.nrao.edu/aipsmemo.html}{http://www.aips.nrao.edu/aipsmemo.html}) following standard procedures. Observations in the A-array configuration were carried out with a bandwidth of 50\,MHz and with the pointing centre at RA: 23\fhr03\fmin03\fs0, Dec: $-$18\degr41\arcmin25\farcs87 (J2000.0) and a total integration time of 168 (spread out on 12 scans) and 46 min (spread out on 5 scans), respectively. For flux and the phase calibration, source 3C\,286 and the source J2246$-$121 were used, respectively. Observations in the C-array configuration were carried out with a bandwidth of 50\,MHz. The observations were performed with the same pointing as mentioned above with a total integration time of 10 min. For the flux and the phase calibration, the source 3C\,48 and the source J0202$-$172 were used respectively. After the initial data reduction the two L-band data sets were merged using the task DBCON. After preliminary CLEANing of the L- and C-band maps with the routine {\sc imagr}, several self-calibrations were performed to improve their quality. The two resultant maps were corrected for primary beam attenuation and the images were finally convolved with circular Gaussians. The flux density was on the \citet{1977A&A....61...99B} scale. The resulting maps are presented in Fig.~\ref{fig3}. Details about the observations and resultant maps is given in Table~\ref{tabnew}.

\subsubsection{VLBA observations}
\label{2.1.3}

To study the nuclear jet kinematics of PKS\,2300$-$18, the archival multiepoch X-band data observed with the Very Long Baseline Array (VLBA) and published by \citet{2002ApJS..141...13B},  \citet{2015ApJS..217....4S} and \citet{2016AJ....151..154G} were analysed. The five epoch maps i.e. 1997.499, 1997.652, 2014.411, 2015.441, and 2017.222 used in our analysis were acquired from the Astrogeo Very Long Baseline Interferometry (VLBI) Flexible Image Transport System (FITS) image data base (\href{https://astrogeo.org/vlbi_images/} {https://astrogeo.org/vlbi$_{-}$images/}). The synthesised beam size of all these maps is similar and varies from 1.94 mas $\times$ 0.83 mas to 2.75 mas $\times$ 1.03 mas and the maps' pixel size is 0.2 mas $\times$ 0.2 mas. Details of our analysis are given in Section~\ref{sec3.2} and about the observations and resultant maps are given in Table~\ref{tabnew}.

\begin{table*}
    \centering
    \caption{ Details of archival and survey data of PKS 2300−18 analysed in this work.}
    \resizebox{\textwidth}{!}{ 
    \begin{tabular}{ccccccccc}
        \hline
        Telescope/Survey & Frequency range & Central freq & Proposal code & Date of observations & TOS & Resolution & PA & rms \\
         &  (MHz) &  (MHz) &  & &  (mins) &  (arcsec x arcsec) &  (\degree) &  (mJy beam$^{-1}$)  \\
         (1) & (2) & (3) & (4) & (5) & (6) & (7) & (8) & (9) \\
        \hline
         ASKAP/RACS-low & 744$-$1032 & 888 & AS110 & 2019 Apr - 2020 Jun & 15  & 25 x 25 & $-$ & 0.352 \\
        ASKAP/RACS-mid & 1224$-$1512 & 1368 & AS110 & 2021 Jan 21 & 15  & 8.80 x 7.40 & 72.03 & 0.181 \\
         VLA-D/NVSS & 1344$-$1456$^a$ & 1400 & AC308 & 1993 Sep 20 & 0.7 & 45 x 45 & $-$ & 0.45 \\
        VLA-A & 1440$-$1490 & 1465 & COND & 1982 Jun 25 & 168  & 2.25 x 1.38$^b$ & 0.05$^b$ & 0.033$^b$ \\
        VLA-C & 1440$-$1490 & 1465 & MITC & 1981 Dec 15 & 10  & 2.25 x 1.38$^b$ & 0.05$^b$ &  0.033$^b$ \\
        JVLA-B/VLASS & 2000$-$4000 & 3000 & VLASS & 2019 Jul 06 & $-$ & 2.90 x 1.77 & 56.29 & 0.158 \\
        VLA-A & 4860$-$4910 & 4885 & COND & 1982 Jun 25 & 46  & 0.7 x 0.7 & $-$ & 0.059 \\
        VLBA & 7392$-$7840 & 7624 & BS241 & 2015 Jun 11 & 1.9 & 0.0028 x 0.0010 & 2.51 & 0.15 \\
        VLBA & 8155$-$8555 & 8340 & BB023 & 1997 Jul 02 & 5.4 & 0.0024 x 0.0009 & -7.9 & 0.374 \\
        VLBA & 8155$-$8555 & 8340 & BB023 & 1997 Aug 27 & 5.4 & 0.0022 x 0.0010 & -9.09 & 0.614 \\
        VLBA & 8428$-$8876 & 8668 & UF001 & 2017 Mar 23 & 4.9 & 0.0019 x 0.0008 & -3.46 & 0.178 \\
        VLBA & 8444$-$8892 & 8668 & BG219 & 2014 May 31 & 2.5 & 0.0021 x 0.0008 & -1.32 & 0.171 \\

        \hline
    \end{tabular}
    }
    \label{tabnew}
    \begin{tablenotes}\footnotesize
    \item (a) NVSS frequency ranges: 1344-1386 MHz and 1414-1456 MHz, (b) values for the merged A and C configuration data.
      \end{tablenotes}
\end{table*}

\subsubsection{Data extracted from surveys}
\label{sec2.1.4}

This work made use of the Rapid Australian Square Kilometre Array Pathfinder (ASKAP) Continuum Survey (RACS: \citealt{2021PASA...38...58H}) for low-frequency
radio data analysis. RACS-low is a large sky survey using ASKAP, covering the sky south of declination +41$\degree$ in the low band at a central frequency of 887.5 MHz. The images are convolved to a common resolution of 25 arcsec with a rms sensitivity of 0.25–0.3 mJy \( \text{beam}^{-1} \). 

High-frequency analysis was conducted using the data from the NRAO VLA Sky Survey (NVSS; \citealt{1998AJ....115.1693C}) RACS-mid (\citealt{2023PASA...40...34D}) and VLASS (\citealt{2020PASP..132c5001L}. NVSS is a continuum survey that covers the entire sky north of the -40$\degree$ declination at 1.4 GHz with a resolution of 45\arcsec. The rms brightness fluctuations in total power are about 0.45 mJy \( \text{beam}^{-1} \)  (Stokes I). The RACS-mid is a large sky survey using ASKAP that comprises images covering the whole sky south of Dec +49$\degree$ in the mid band at 1367.5 MHz. The rms sensitivity is $\sim$0.15–0.4 mJy \( \text{beam}^{-1} \) at a resolution of 8.1--47.5 arcsec. VLASS is an all-sky survey above declination -40$\degree$. It operates within a range of 2--4 GHz with an angular resolution of ~2.5 arcsec. Details about the observations and resultant maps are given in Table~\ref{tabnew}. Throughout this paper, the flux-density errors were calculated using the following formula:

\begin{equation}
    \Delta S = \sqrt{(S \times \Delta S_c)^2 + \left( rms \times \sqrt{\frac{area}{beam}} \right)^2}
    \label{eq:placeholder}
\end{equation}

where $\Delta S_c$ is the calibration error taken as 10 per cent for the uGMRT observations and 5 per cent for the VLA observations. The calibration error for the radio surveys was taken as 5 per cent.

\subsection{Optical observations}
\label{sec2.2}

Spectral observations of PKS\,2300$-$18 were obtained from the Isaac Newton Group of Telescopes archives. The target was observed using the William Herschel Telescope (WHT) and the Intermediate-dispersion Spectrograph and Imaging System (ISIS). The observations were taken on  July 4 1994, with gratings R158B and R158R, slit width 1\farcs34, and 900 sec of exposure. During the observations, seeing (0\farcs9) was smaller than the slit width, ensuring that none of the quasar emission was missed. The spectrum was taken at two position angles (measured from north to east), 167\degree\,and 162\degree; however none of these positions reached the companion galaxy.
The spectrum was wavelength calibrated using CuNe+CuAr arc lamp exposures, and photometrically calibrated using exposures of Feige 34 and Feige 92 standard spectrophotometric stars taken on the same night.\\

\subsection{X-ray observations}
\label{sec2.3}

To investigate the properties of the hot gas associated with the host galaxy of PKS\,2300$-$18, as well as its AGN, we used archival X-ray data from the ROentgen SATellite \citep[{\sc ROSAT};][]{truemper82} All Sky Survey \citep[{\sc RASS}:][]{1999A&A...349..389V} and from the Chandra X-ray Observatory Data Archive.\footnote{https://cxc.harvard.edu/cda/}

\subsubsection{ROSAT data}
\label{sec2.3.1}

In this work we used the broad-band image from the sequence rs932061n00, smoothed to the resolution of 2 arcmin, which is the PSF of the Position Sensitive Proportional Counter (PSPC) instrument. The resulting image was used only to estimate the extent of the X-ray emission associated with our studied source, which will be discussed in Section ~\ref{sec6.4.2}.

\subsubsection{Chandra data}
\label{sec2.3.2}

PKS\,2300$-$18 was observed with Chandra on September 11 2021 for a total of 9.57\,ks, using an ACIS-S camera. In this work, we use the pipeline data obtained from the Chandra Data Archive. However, the data were reprocessed with the $chandra-repro$ script for the most accurate and up-to-date calibration. Due to the high brightness of the source associated with the  PKS\,2300$-$18 host galaxy, the data included a readout streak, which was removed using the $acisreadcorr$ script. Next, we produced a broad-band (0.2--7 keV) image and smoothed it with the PSF of the instrument. The extended emission visible in the ROSAT map (left insert in Fig.~\ref{fig1}) was not detected, likely due to short time of the observations and overall lower sensitivity to diffuse structures of the Chandra telescope. A circular region of 4 arcsec radius around the brightest emission associated with the host galaxy of PKS\,2300$-$18 was used to extract the spectrum. The background spectrum was also extracted from a rectangular source-free area close to the studied source. Further, for both spectra, response files were created. The final background-subtracted source spectrum was grouped (25 counts in an energy bin) to increase the signal-to-noise ratio. The spectrum of PKS\,2300$-$18 was analysed with XSPEC12 \citep{arnaud96}. The results of the model fitting are presented in Section~\ref{sec5}.

\section{Multifrequency radio analysis}
\label{sec3}

\subsection{Comprehensive radio morphology}
\label{sec3.1}

The source displays a pair of S-shaped, inversion-symmetric jets oriented along the north-south direction, extending over 240 kpc, as observed in the low-frequency uGMRT maps shown in Fig.~\ref{fig2} and the high-frequency VLA maps shown in Fig.~\ref{fig3}. These jets are embedded within low surface brightness diffuse emission that extends along the southwest and northeast direction, forming a structure of total length $\sim$5 arcmin. The S-shaped jets resemble an FRI-type morphology with $P_{\rm1.4 GHz} \simeq 5.5 \times 10^{25}{\rm W\, Hz^{-1}}$, calculated using a spectral index value of 0.7 and total flux density measured at 1.4 GHz NVSS of 1.38 Jy, using equation (1) of \citet{2001AJ....121.2381B}.
 The high-resolution L-band map at 1.4 GHz and C-band map at 6 GHz in the A configuration seen in Fig.~\ref{fig3} have a well-resolved view of the core and the jets, where the radio core is observed to have enhanced emission at higher frequencies. The core also coincides perfectly with the position of the optical host at the centre. In Fig.~\ref{fig3}, it can be observed that the northern jet is pronounced and well collimated compared with the southern jet. The northern jet features three distinct knots labelled K1, K2, and K3, with K1 being the brightest, followed by K2 and K3; their flux density values are given in Table~\ref{tab2}. The southern jet is much fainter and quite diffuse in comparision with the northern jet, and lacks any complementary knot-like features. This could most likely result from relativistic Doppler beaming of the jets.  \\

The diffuse wings are aligned in the northeast and southwest directions relative to the central S-shaped jets. Notably, the upper part of the northeastern wing features a prominent protrusion that resembles an inverted U, as clearly depicted in the uGMRT 400 MHz map (Fig.~\ref{fig2}). A subtler complementary protrusion is observed in the lower part of the southwestern wing near its bottom edge. This pattern suggests an earlier orientation of the jets, indicating that the current structure is the result of a jet reorientation. From the current inversion-symmetric S-shaped jets we can hence deduce that the source is going through precession. The overall morphology of the source can be attributed to jet reorientation taking place in the counter-clockwise direction, due to a continuous jet precession. The jet precession not only occurs in a counterclockwise direction but also involves the northern jet approaching our line of sight, causing Doppler boosting of its components, while the southern jet is receding. This may explain the presence of knots in the northern jet and their absence in the southern jet. The total extent of the source, taking into account the extended wings, is $\sim$760 kpc in linear size.

\begin{figure*}
    \centering
    \includegraphics[width=1\linewidth]{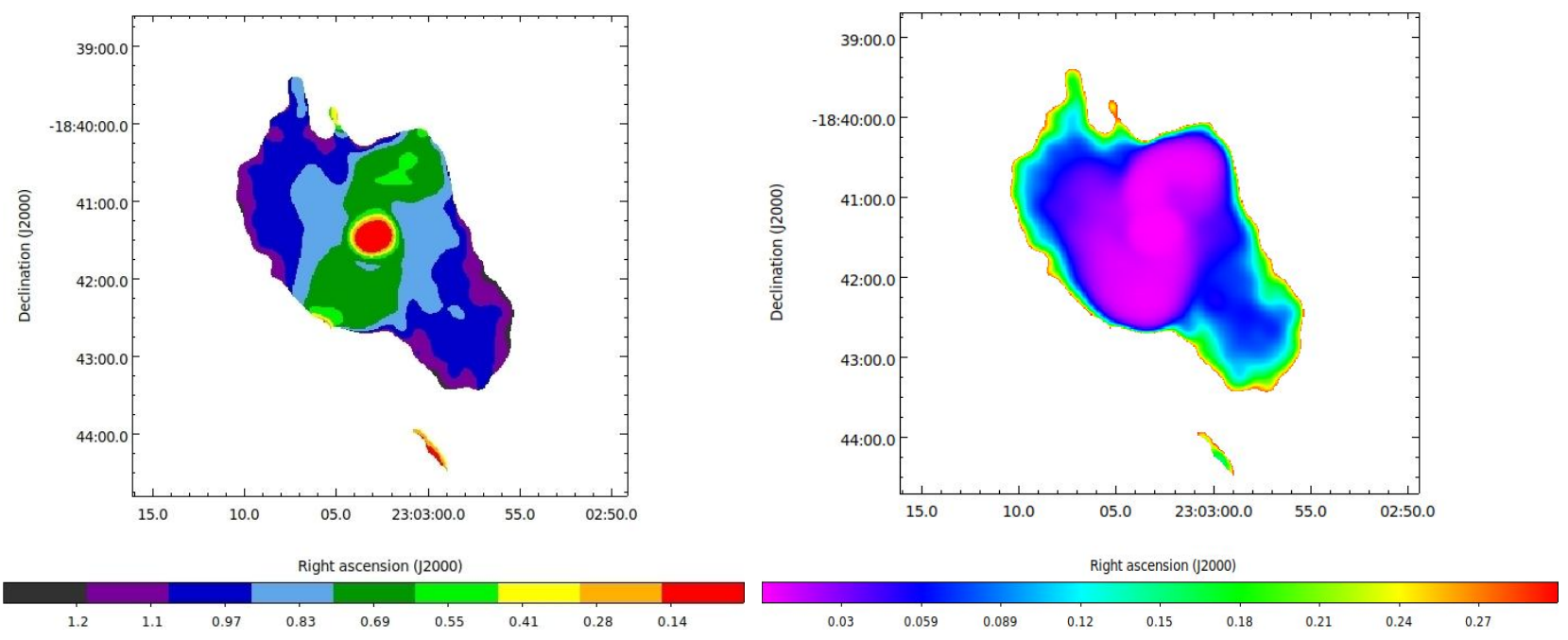}
    \caption{Left: The spectral index map of PKS\,2300$-$18 between GMRT 400 MHz and JVLA 6 GHz. The colour bar shows the spectral index values. Right: The spectral index error map between GMRT 400 MHz and JVLA 6 GHz with the colour bar showing error associated with the spectral index values.}
    \label{fig7}
\end{figure*}

\subsubsection{Spectral index analysis}
\label{sec3.1.1}

The evolution of plasma in the radio lobes as a result of synchrotron and inverse-Compton losses is well traced on the spectral index (SI) map. The SI map for PKS\,2300$-$18 was produced using the uGMRT band 3 (400 MHz) and the large-scale low-resolution JVLA C-band (6 GHz) maps with similar uv coverage. These maps were convolved to a resolution of 15 arcsec $\times$ 15 arcsec, which is comparable with the original resolution of the uGMRT and the JVLA maps. AIPS task {\sc hgeom} and {\sc comb} were subsequently used to align the geometry of the maps and masking of three times above the root- mean-square noise level was applied to produce the SI map shown in Fig.~\ref{fig7}. The SI map ($S_{\nu} \propto \nu^{-\alpha}$) distribution displays remarkable variation within the source and shows a smooth gradient that runs from the core to the extended wing regions. The S-shaped jets emerging from the flat-spectrum radio core are distinctively visible on the SI map and have a uniform $\alpha$$\sim$ 0.7. The small region on either side of the jets, with comparatively flatter spectra, could be due to in situ reacceleration of particles (\citealt{2024JApA...45...12G}) where the jet experiences bending. The extended wings surrounding the jet and the core have steeper spectra and show an evolution in their SI, with the steepest plasma being present at the edge of the wings. The low surface brightness wings have steeper spectra compared with the central S-shaped structure. The gradient in the SI map traces the location of the oldest to the youngest plasma, tracking the motion of the radio jets over a period of time. 

The current trend in the SI map hints at an anticlockwise movement of the radio jet, moving from the extended wings to the S-shaped jets, with the north jet bending towards the western side and the south jet bending towards the eastern side, forming an inversion-symmetric S-shape. This implies a previous southwest and northeast alignment of the radio jets, where we currently see steep spectra. The jets gradually moved in the anticlockwise direction to their current north-south alignment, where we now see comparatively flatter spectra. The SI layout of this source shares similarities with the X-shaped radio source J0113+0106 (\citealt{2019AJ....157..195L}), diverging from the typical radio lobes characteristic of classical FRII or FRI sources. This unusual morphology prompts several questions about the formation, dynamics, and evolution of the source. Hence, the SI pattern and structure of PKS\,2300$-$18 will provide valuable insights into the nature of precessing jet radio galaxies in general.

\begin{table}
\caption{Flux-density values of radio knots at different frequencies given in mJy (see Fig.~\ref{fig3}).}
    \centering
    \begin{tabular}{ccccc}
    \hline
        Region & 1367 MHz  & 1464 MHz  & 3000 MHz  & 4885 MHz\\
        \hline
        K1 & 72.9$\pm$6.3 & 32.5$\pm$1.6 & 15.5$\pm$2.4 & 11.2$\pm$0.6\\
        K2 & 77.8$\pm$3.9 & 53.4$\pm$2.7 & 33.9$\pm$ 5.1 & 16.0$\pm$0.8  \\
        K3 & 59.4$\pm$3.0 & 50.0$\pm$ 2.4 & 29.7$\pm$4.9 & 11.5$\pm$0.6   \\
    \hline
    \end{tabular}
    \label{tab2}
    \begin{tablenotes}\footnotesize
\item The flux density values of column (2)--(5) are from RACS survey, archival VLA observations, VLASS survey, and archival VLA observations, respectively, as detailed in Section~\ref{sec2}.
\end{tablenotes}
\end{table}

\subsubsection{Polarisation}
\label{sec3.1.2}

\begin{figure}
 
  \includegraphics[scale=0.70]{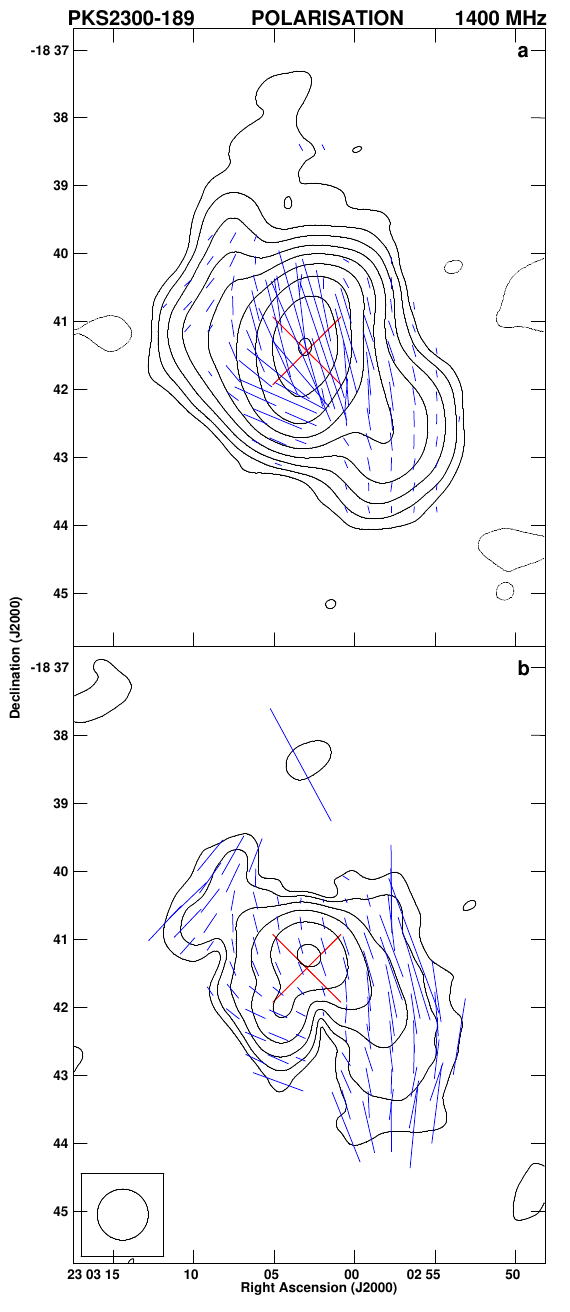}
  \caption{ 1400 MHz NVSS polarimetric images. (a) Total intensity contours spaced by a factor of $\sqrt2^{n}$ starting at 1.8 mJy beam$^{-1}$. Superimposed are E-vectors rotated by 90$\degr$ with their lengths proportional to the polarised intensity, where 1 arcmin corresponds to a 12 mJy beam$^{-1}$. (b) Linearly polarised intensity contours spaced by a factor of $\sqrt2^{n}$ starting with the 0.9 mJy beam$^{-1}$ with the vectors of the fractional linear polarisation superimposed. A length of 1 arcmin corresponds to 30\% of the fractional linear polarisation. The red X-sign marks the central position of the host galaxy. The size of the convolved beam is indicated by a circle in the bottom-left corner of the image. }
  \label{fig9}
\end{figure}

Maps of the linearly polarised intensity and fractional polarisation were created by combining the NVSS Stokes Q and U maps using the AIPS task {\sc comb}. These maps allow for the determination of the polarised flux density, fractional polarisation, and the polarisation angle of the E-vector. The total intensity NVSS map with the electric field E-vectors (rotated by 90\degr) superimposed is shown in Fig.~\ref{fig9}(a) and Fig.~\ref{fig9}(b) shows the linearly polarised intensity map with the vectors of fractional linear polarisation superimposed.  
It is observed that the whole structure is polarised. The total integrated polarised flux intensity of this source is $96.2\pm5.1$ mJy,
which gives $m=6.9$ per cent for the mean fractional polarisation (m is the ratio of polarised flux to total flux). \citet{1981ApJS...46..239S} presented the polarisation study at 2695 and 8085 MHz, where the fractional polarisation
was found to be m = 5.3 ± 0.5 per cent and m = 2.0±1.2 per cent, respectively. There is a clear trend here showing a monotonic decrease in the degree of polarisation with the increasing wavelength.
As anticipated, steep-spectrum radio galaxies typically exhibit a reduction in the degree of polarisation at lower frequencies.
(e.g., \citealt{1988ARA&A..26...93S}; \citealt{2003A&A...406..579K}).
However, \citet{1981ApJS...45...97S} have shown (in their Figure 1) some exceptions where the degree of polarisation of a source first increases and then decreases with increasing wavelength. The authors explain such a behavior due to the existence of two or more polarised components with different rotation measures within the source. Another possible reason
could be connected to the spectral index differences between less- and more-polarised components. To precisely determine the phenomenon taking place in PKS\,2300$-$18, additional sensitive polarimetric observations are necessary. In the NVSS maps most of the polarised flux is concentrated in the central part of the source. However, fractional polarisation is larger in the diffuse wings. The magnetic field vectors in the central part of the source, which contain the core and the S-shaped jets, are aligned in the north-south direction. The magnetic field vectors also follow the direction of the jet path closely in the diffuse wings, which is traced by the S-shaped jets. 
\citet{1999ApJS..124..285R} showed a high-resolution ($\sim$3 arcsec) map of the central part of this source at 4735.1 MHz where
the core, the northern blobs, and the S-structure were visibly polarised.

\subsection{VLBI pc-scale jets}
\label{sec3.2}

\begin{figure}
    \centering
    \includegraphics[width=1.08\linewidth]{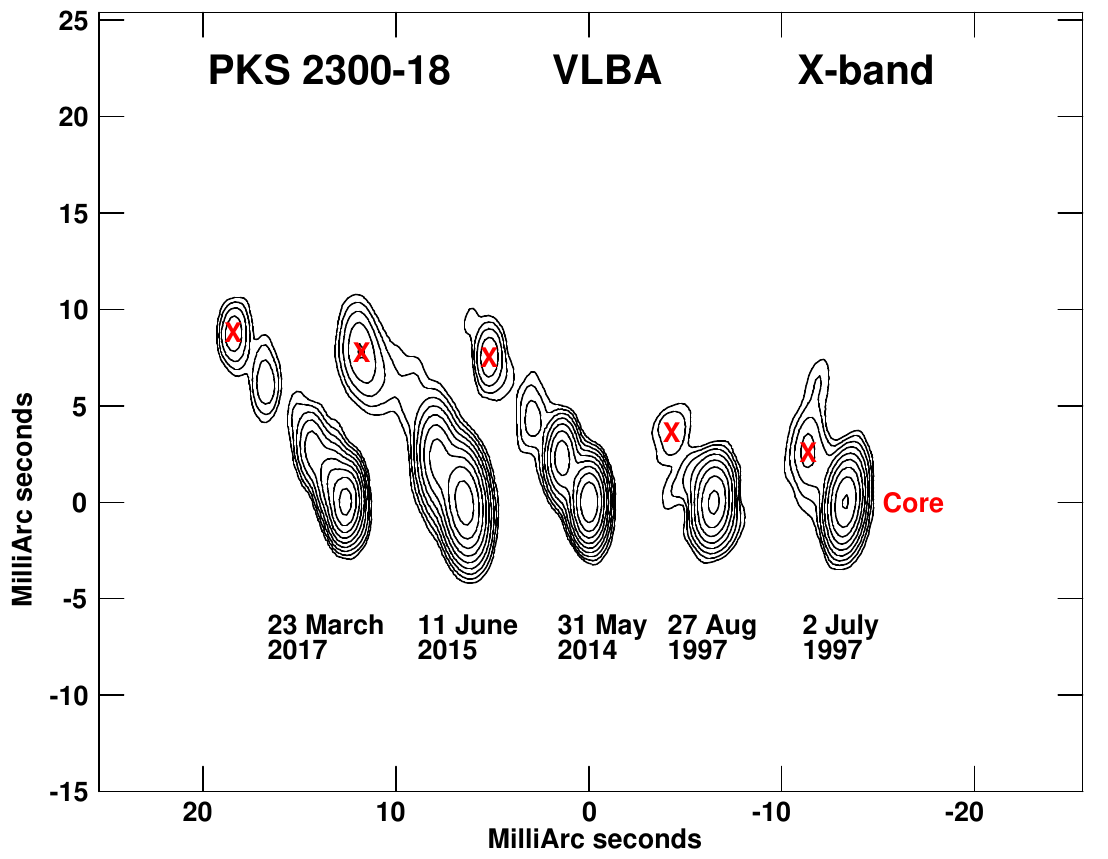}
    \caption{Apparent superluminal motion in the nucleus of the quasar PKS\,2300$-$18. VLBA X-band five-epoch images of the variable parsec-scale structure of the jet are shown.  Positions of the core and a bright component named as `X' blob at the forefront of the jet are marked in red.}
    \label{fig5}
\end{figure}

The milliarcsecond radio structure from VLBA observations of the central part of PKS\,2300$-$18 is shown in Fig.~\ref{fig5}. The brightness distribution reveals a compact core and an extended and weaker one-sided jet component of more than 10 mas length in the north-east direction of the core. The jet exhibits a multi-blob structure.
A prominent component located at the head of the jet, visible in all epochs is marked with a red `X' sign in Fig. ~\ref{fig5}.
The exact positions of the core and the `X' blob were determined using the AIPS task {\sc jmfit}. The position angle of the `X' blob with respect to the core in the epoch 2017.222 is about 34\degree. For comparison, the position angle of K1 is 20\degree.
We estimated the apparent proper motion by tracking the positional changes in the `X' component with respect to the core, identified across five different epochs covering a period of 20 years.  
According to e.g. \citet{2012ApJ...760...77A}, for example,
the apparent transverse velocity $v_{app}$ in the rest frame of the source,
in units of $c$ can be obtained as: $v_{app}=0.0158 \mu D_A (1+z)$,
where $D_A$ is the angular size distance of the source in Mpc
and $\mu$ is the angular separation velocity of the `X' blob with respect to the central core.
$D_A$ to PKS\,2300$-$18 is 472.6 Mpc, while 
$\mu$ is in mas yr$^{-1}$ and is defined as $\mu={\theta_{X-Core}}/{\Delta t}$.
The angular separation velocity and apparent proper motion obtained in the source rest frame of X are $\mu=0.3486\pm0.0158$ mas yr$^{-1}$ and  
$v_{app}=(2.31+/- 0.10$) c.
The kinematical age in the source rest frame, calculated from ${\theta_{X-Core}}/{\mu (1+z)}$, is 26.1 yr.\\

\begin{figure}
    \centering
    \includegraphics[width=1.12\linewidth]{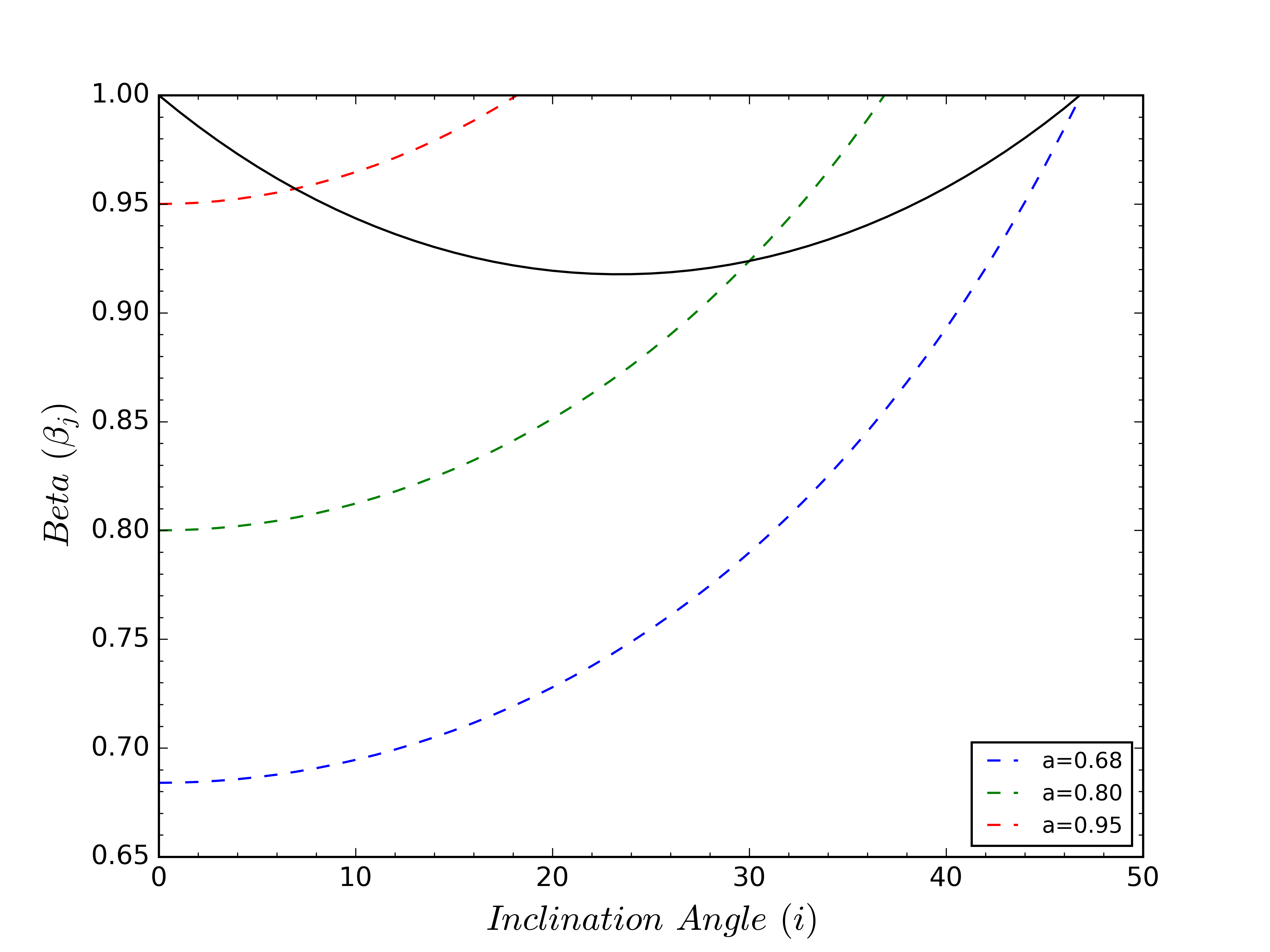}
    \caption{The dependence of the jet velocity $\beta_j$ on the pc-scale jet inclination angle, i in degrees, for PKS\,2300$-$18.
The black solid line represents the equation (2) and the dashed lines
represent the equation (3) for three different values of a: 0.95 (red), 0.8 (green), and 0.684 (blue) as detailed in the Section~\ref{sec3.2}.}
    \label{fig6}
\end{figure}

The jets of PKS\,2300$-$18 show mildly superluminal speeds at 2.3c on the parsec scale. Such superluminal motions in AGNs were predicted by \citet{1966Natur.211..468R}. The first detection of a source moving with an apparent velocity higher than the speed of light was the quasar 3C\,279 (\citealt{1969Natur.224.1094G}). To date, apparent speeds observed in AGNs range from subluminal speeds to a maximum of about 50c (QSO PKS\,J0808$-$0751: \citealt{2009AJ....138.1874L}), although common values are 1--10c. Assuming intrinsically symmetric jets and having the apparent proper motion in the source rest frame calculated above, following e.g. \citet{1993ApJ...407...65G} we estimated the inclination angle of the pc-scale jet, i, with respect to the observer’s line of sight as
\begin{equation}
v_{app}= \left[c \beta_j \sin(i)\right]/\left[1 - \beta_j \cos 
(i)\right], 
\end{equation}
where $\beta_j$ is the velocity of the jet in units of c. 
Using the above value of $v_{app}$, we can write the above equation as
\begin{equation}
\beta_j=2.31/\left[\sin(i)+2.31\cos(i)\right],
\end{equation}
further, following, e.g. \citet{2010A&A...523A...9H},
\begin{equation}
\rm \beta_j = a/\cos(i), 
\end{equation}

where $\rm a =  \left[s-1\right]/\left[s+1\right]$, $\rm s = J^{1/(2+\alpha)}$, 
J is the jet-to-counter-jet flux-density ratio, and {$\alpha$} is the spectral index of the jet.
The graphical solution of the above two equations for $\beta_j$ and i for three values of a= 0.95, 0.8, and 0.684 is shown in Fig. ~\ref{fig6}. The intersection of the solid black line with the three different dashed lines gives a solution to the above equations. Therefore, we infer that the current angle to the line of sight of PKS\,2300$-$18 cannot be larger than 46\degree.9, and this is valid for a=0.684 (when $\beta_j=1$). The above limit value of a corresponds to J=65.6 (for $\alpha=0.5$). From the X-band epoch 2017.222 map, we obtained the `X' blob peak flux-density of $8.3\pm0.17$ mJy  beam$^{-1}$, and since there was no counter-blob visible on the map where the rms value is 0.2 mJy beam$^{-1}$, the lower limit of J was calculated as $\sim$ 41.7.
\\

\begin{figure}
    \centering
    \includegraphics[width=1\linewidth]{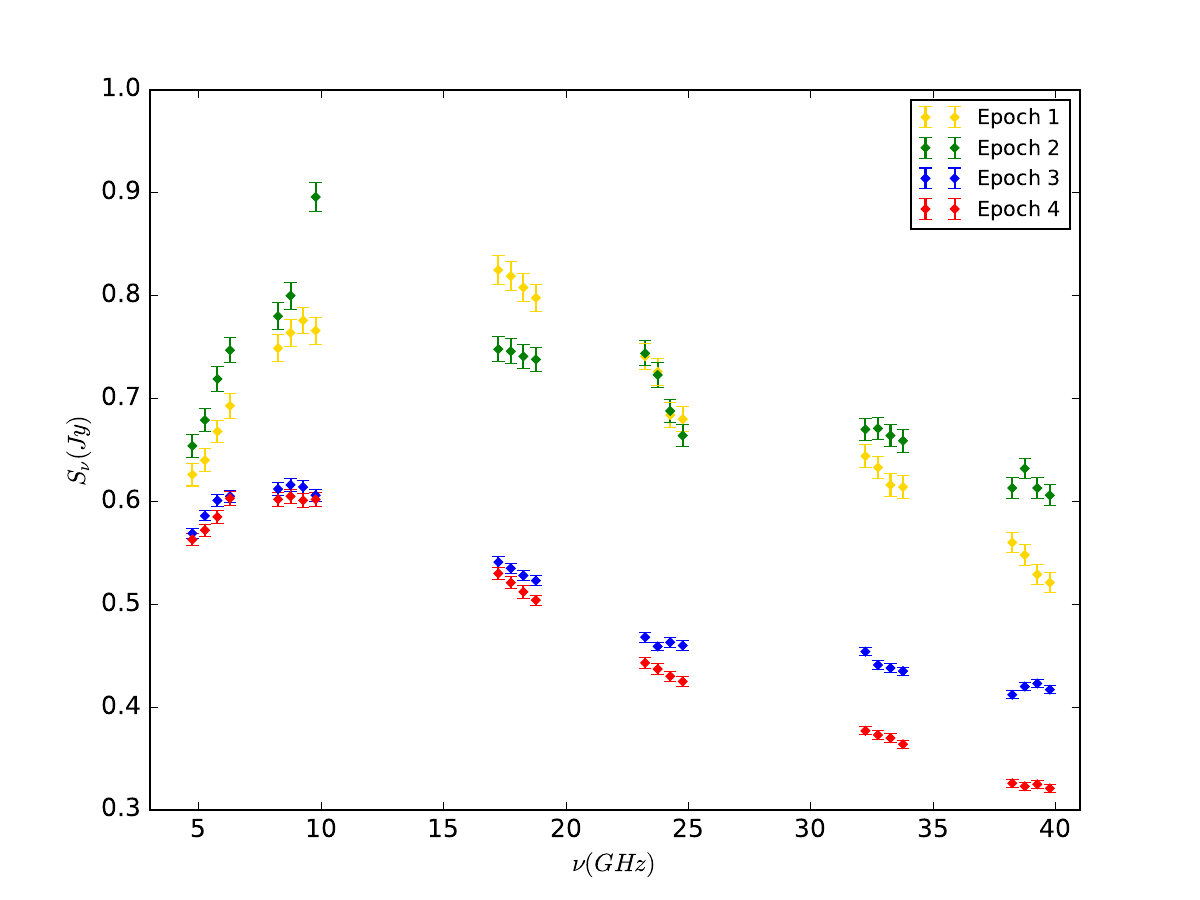}
    \caption{Flux-density values of the radio core of PKS\,2300$-$18 in the frequency range 5-40 GHz showing variability (details in Section~\ref{sec3.3}).}
    \label{fig4}
\end{figure}

\subsection{Radio core}
\label{sec3.3}

The radio core of  PKS\,2300$-$18 was first observed by \citet{1980ApJS...44..151T} in 1979 using the VLA at 4885 MHz, and the flux density was found to be 342$\pm$14 mJy. Concurrently, the core flux density was measured by \citet{1984MNRAS.207...55H} between 1978--1979 as $287\pm30$ mJy at 2695 MHz and $ 304\pm45$ mJy at 8085 MHz. Follow-up observations in 1982 using the VLA in the A configuration revealed a core flux density of $253\pm8$ mJy at 1465 MHz and $417\pm17$ mJy at 4885 MHz. This measured flux at 5 GHz was significantly higher than the value obtained in 1979. This indicated a flat-spectrum radio core at higher frequencies, which usually implies synchrotron self-absorption. The radio core at higher frequencies also shows an inverted spectrum, which influences the total flux of the source at higher frequencies.

The source was reobserved to study spectral behaviour and variability between 5 and 40 GHz as part of a sample of galaxies for the PACO project in several epochs between 2009 and 2010. The epoch 1 observations were conducted on 2009 November 11, epoch 2 on 2009 December 4, epoch 3 on 2010 June 14, and epoch 4 on 2010 June 19. The flux-density values of all four epochs are plotted in Fig.~\ref{fig4}. The core spectrum shows a consistent pattern throughout the four epochs, with the core displaying an inverted spectrum up to 10 GHz, followed by a steepening of the spectrum up to 40 GHz. The observations between epoch 1 and epoch 2, which were taken a few weeks apart, reveal considerable variations in the flux density values across the observed spectral range. A similar pattern is observed between epochs 3 and 4, with these observations spaced less than a week apart. There are noticeable variations in the flux density values, particularly at higher frequencies from 22 GHz upwards, where prominent changes are observed between epochs.

The variability of radio quasars is particularly noticed and studied in the case of blazars such as 3C 466 (\citealt{1988ApJ...331..746B}), OJ 287 (\citealt{2023ApJ...944..177K}), and S5 0716+714 (\citealt{2013A&A...552A..11R}). There are several factors that can cause such variability, including Doppler boosting in blazars or jets with a low inclination angle to the line of sight. Additionally, relativistic beaming can also lead to anisotropic emission, as noted by \citet{1966ApJ...146..597W}. Extrinsic factors often include interstellar scintillations that can cause flux-density variations on time scales of a few hours or days (\citealt{2018MNRAS.474.4396K}). In the case of PKS\,2300$-$18, the observed variations occur over periods ranging from a few weeks to several years, with minimal indications of fluctuation during these observation windows. A possible cause for this variability could also be inherent to the source itself, as suggested by \citet{1995ARA&A..33..163W}, potentially arising from (a) a change in the accretion rate due to disc instabilities and thus in jet power and brightness, (b) propagation of shocks along the jets, or (c) precessing jets that could vary the Doppler beaming of the ejected jet components.
As the core variability seen in the case of PKS\,2300$-$18 is on the timescale of few days to months, it is likely to have a point of origin very close to the core, possibly either due to some variation in the accretion rate or propagation of shocks in the pc-scale jets. However, other scenarios might also contribute simultaneously towards it (\citealt{2023NatAs...7.1282R}).

\begin{figure}
    \centering
    \includegraphics[width=0.8\linewidth]{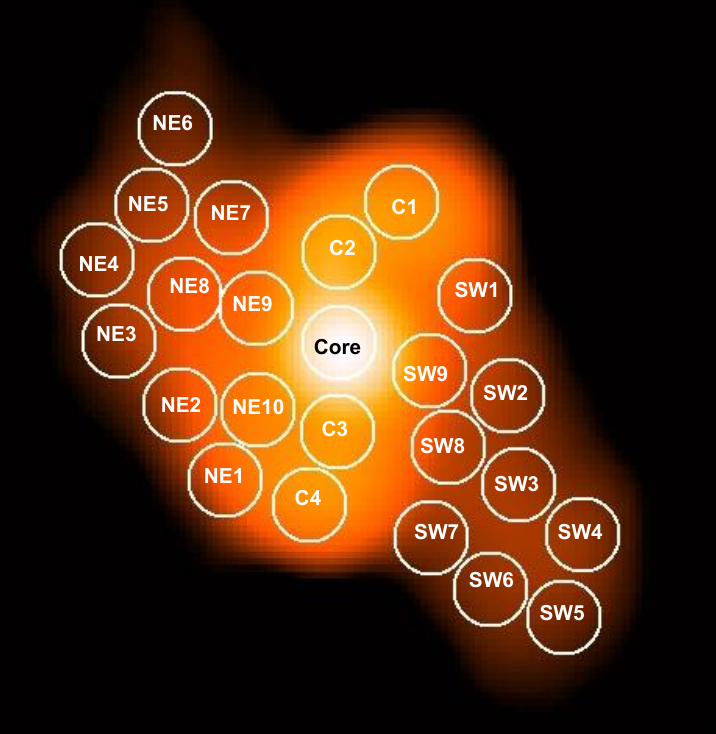}
    \caption{JVLA 6 GHz map displaying different regions of the source selected for fitting particle injection model and ageing analysis. }
    \label{fig8}
\end{figure}

\begin{table*}
\caption{Flux-density values (mJy) of different segmented regions of PKS\,2300$-$18.}
\label{tab:landscape2}
\begin{tabular}{ccccccccc}

\hline

  Region/Freq. & 183 MHz & 322 MHz & 400 MHz & 607 MHz & 888 MHz & 1367 MHz & 6000 MHz \\
  (1)  & (2) & (3) & (4) & (5)& (6) & (7) & (8)  \\
 \hline
 \\
 Total  & 4638$\pm$464 & 3325$\pm$332 & 2929$\pm$293 & 2301$\pm$161 & 1976$\pm$99 & 1364$\pm$68  & 805$\pm$40\\
NE1  & 8.7$\pm$2.1 & 7.0$\pm$0.7 & 5.6$\pm$0.6 & 4.0$\pm$0.4 & 3.5$\pm$0.4 & 3.0$\pm$0.4  & 0.9$\pm$0.1\\
NE2 & 8.5$\pm$2.1 & 6.6$\pm$0.7 & 5.7$\pm$0.6 & 3.6$\pm$0.4 & 3.2$\pm$0.4 & 2.6$\pm$0.4  & 0.5$\pm$0.1\\
NE3  & 4.7$\pm$2.0 & 2.9$\pm$0.4 & 3.2$\pm$0.4 & 2.3$\pm$0.4 & 1.9$\pm$0.4 & 1.3$\pm$0.4  & 0.2$\pm$0.1\\
NE4  & 4.5$\pm$1.9 & 4.1$\pm$0.5 & 3.4$\pm$0.4 & 2.5$\pm$0.4 & 1.8$\pm$0.4 & 1.3$\pm$0.4  & 0.2$\pm$0.1\\
NE5  & 4.5$\pm$1.9 & 4.9$\pm$0.5 & 3.9$\pm$0.4 & 2.5$\pm$0.4 & 2.2$\pm$0.4 & 1.3$\pm$0.4 & 0.3$\pm$0.1\\
NE6  & 2.9$\pm$1.9 & 3.3$\pm$0.4 & 2.5$\pm$0.3 & 1.5$\pm$0.4 & 1.4$\pm$0.4 & 0.8$\pm$0.4  & 0.2$\pm$0.1\\
NE7  & 5.9$\pm$2.0 & 5.3$\pm$0.6 & 5.0$\pm$0.5 & 3.0$\pm$0.4 & 2.7$\pm$0.4 & 1.9$\pm$0.4  & 0.4$\pm$0.1\\
NE8  & 10.1$\pm$2.1 & 6.4$\pm$0.7 & 6.8$\pm$0.7 & 5.0$\pm$0.5 & 3.9$\pm$0.4 & 2.8$\pm$0.4  & 0.6$\pm$0.1\\
NE9  & 16.3$\pm$2.5 & 10.8$\pm$1.1 & 10.2$\pm$1.0 & 6.9$\pm$0.6 & 6.2$\pm$0.5 & 5.0$\pm$0.5  & 1.5$\pm$0.1\\
NE10  & 26.1$\pm$3.2 & 19.9$\pm$2.0 & 16.1$\pm$1.6 & 11.2$\pm$0.9 & 10.0$\pm$0.6 & 8.1$\pm$0.6  & 2.3$\pm$0.2\\

SW1  & 5.9$\pm$2.0 & 6.1$\pm$0.7 & 4.5$\pm$0.5 & 2.8$\pm$0.4 & 2.9$\pm$0.4 & 2.0$\pm$0.4  & 0.5$\pm$0.1\\
SW2  & 4.2$\pm$1.9 & 2.7$\pm$0.4 & 3.5$\pm$0.4 & 2.5$\pm$0.4 & 2.0$\pm$0.4 & 1.4$\pm$0.4 & 0.3$\pm$0.1\\
SW3  & 5.5$\pm$2.0 & 4.6$\pm$0.5 & 3.6$\pm$0.4 & 2.6$\pm$0.4 & 2.1$\pm$0.4 & 1.2$\pm$0.4  & 0.3$\pm$0.1\\
SW4  & 4.2$\pm$1.9 & 4.2$\pm$0.5 & 3.5$\pm$0.4 & 2.7$\pm$0.4 & 2.2$\pm$0.4 & 1.1$\pm$0.4  & 0.2$\pm$0.1\\
SW5  & 3.8$\pm$1.9 & 2.8$\pm$0.4 & 2.6$\pm$0.3 & 1.6$\pm$0.4 & 1.5$\pm$0.4 & 0.8$\pm$0.4  & 0.2$\pm$0.1\\
SW6  & 3.4$\pm$1.9 & 3.5$\pm$0.4 & 3.0$\pm$0.4 & 2.2$\pm$0.4 & 1.8$\pm$0.4 & 1.0$\pm$0.4&  0.2$\pm$0.1\\
SW7  & 3.8$\pm$1.9 & 4.0$\pm$0.5 & 2.7$\pm$0.3 & 1.5$\pm$0.4 & 1.6$\pm$0.4 & 1.3$\pm$0.4  & 0.2$\pm$0.1\\
SW8  & 6.0$\pm$2.0 & 3.9$\pm$0.5 & 4.2$\pm$0.5 & 3.3$\pm$0.4 & 2.7$\pm$0.4 & 1.8$\pm$0.4  & 0.4$\pm$0.1\\
SW9  & 11.8$\pm$2.2 & 9.9$\pm$1.0 & 7.6$\pm$0.8 & 5.4$\pm$0.5 & 5.6$\pm$0.5 & 3.8$\pm$0.4  & 1.5$\pm$0.1\\
C1  & 34.8$\pm$4.0 & 22.9$\pm$2.3 & 20.6$\pm$2.1 & 13.6$\pm$1.0 & 14.0$\pm$0.8 & 11.2$\pm$0.7  & 3.8$\pm$0.2  \\
C2  & 68.5$\pm$7.1 & 47.1$\pm$4.7 & 42.1$\pm$4.2 & 28.8$\pm$2.0 & 28.0$\pm$1.4 & 22.5$\pm$1.2   & 7.6$\pm$0.4\\
C3  & 48.1$\pm$5.2 & 34.2$\pm$3.4 & 29.8$\pm$3.0 & 20.2$\pm$1.5 & 18.8$\pm$1.0 & 14.3$\pm$0.8  & 5.1$\pm$0.3\\
C4 & 34.1$\pm$3.9 & 23.0$\pm$2.3 & 20.1$\pm$2.0 & 14.2$\pm$1.1 & 12.5$\pm$0.7 &  10.0$\pm$0.6 &  3.3$\pm$0.2\\

\\
\hline

\hline
\end{tabular}
\begin{tablenotes}\footnotesize
\item The flux-density values of column (2)$-$(5) and column (8) are taken from dedicated observations as detailed in Sections~\ref{sec2.1.1} and ~\ref{sec2.1.2} and the flux- density values of column (6) and (7) are taken from the RACS survey, as detailed in Section~\ref{sec2.1.4}. The regions are defined in Fig.~\ref{fig8}.
\end{tablenotes}
\label{tab3}
\end{table*}

\subsection{Analysis of radio spectra of the wings}
\label{sec3.4}

\subsubsection{Source energetics}
\label{sec3.4.1}

Electrons present inside a fixed magnetic field emit radiation primarily via synchrotron processes. At higher frequencies, this leads to higher energy electrons losing energy much faster than their lower energy counterparts. This produces spectra that are increasingly curved over time when there is no further reacceleration of particles. To trace the movement of plasma over time and track the jet reorientation in PKS\,2300$-$18, the north eastern (NE) wing was divided into 10 separate circular regions, and the south western wing (SW) was divided into nine separate circular regions (see Fig.~\ref{fig8}), based on the spectral gradient seen in Fig. ~\ref{fig7}, with the sampling region size similar to the radio beam (28 arcsec).
To calculate the radiative losses over time in these regions, the Jaffe \& Perola \citep[JP:][]{1973A&A....26..423J} model was used to fit the spectrum of PKS\,2300$-$18. All regions of the NE and SW wings were fitted with the JP model due to their steep spectra and the absence of any compact hotspots in the wings. This model uses an initial power-law energy distribution to determine the time evolution of the radiative loss incurred by high-energy electrons as a result of synchrotron and inverse-Compton processes. This results in a higher degree of curvature in the spectrum consisting of older plasma.

Model fitting was carried out by analysing the emission spectrum of the particles characterised by their injection spectral index ($\alpha_{\rm{inj}}$), distributed uniformly in pitch angle with respect to the direction of the magnetic field.
The JP model was applied with the following assumptions: (i) the particles adhere to a constant power-law energy distribution without any reacceleration of radiating particles once they enter the regions, (ii) the magnetic field is constant in time and uniform in space and the field lines are tangled with the field intensity remaining constant during the energy dissipation, (iii) the isotropisation time is small compared with the radiative lifetime for the pitch angles of the injected particles.

\begin{figure}
    \centering
    \includegraphics[width=1\linewidth]{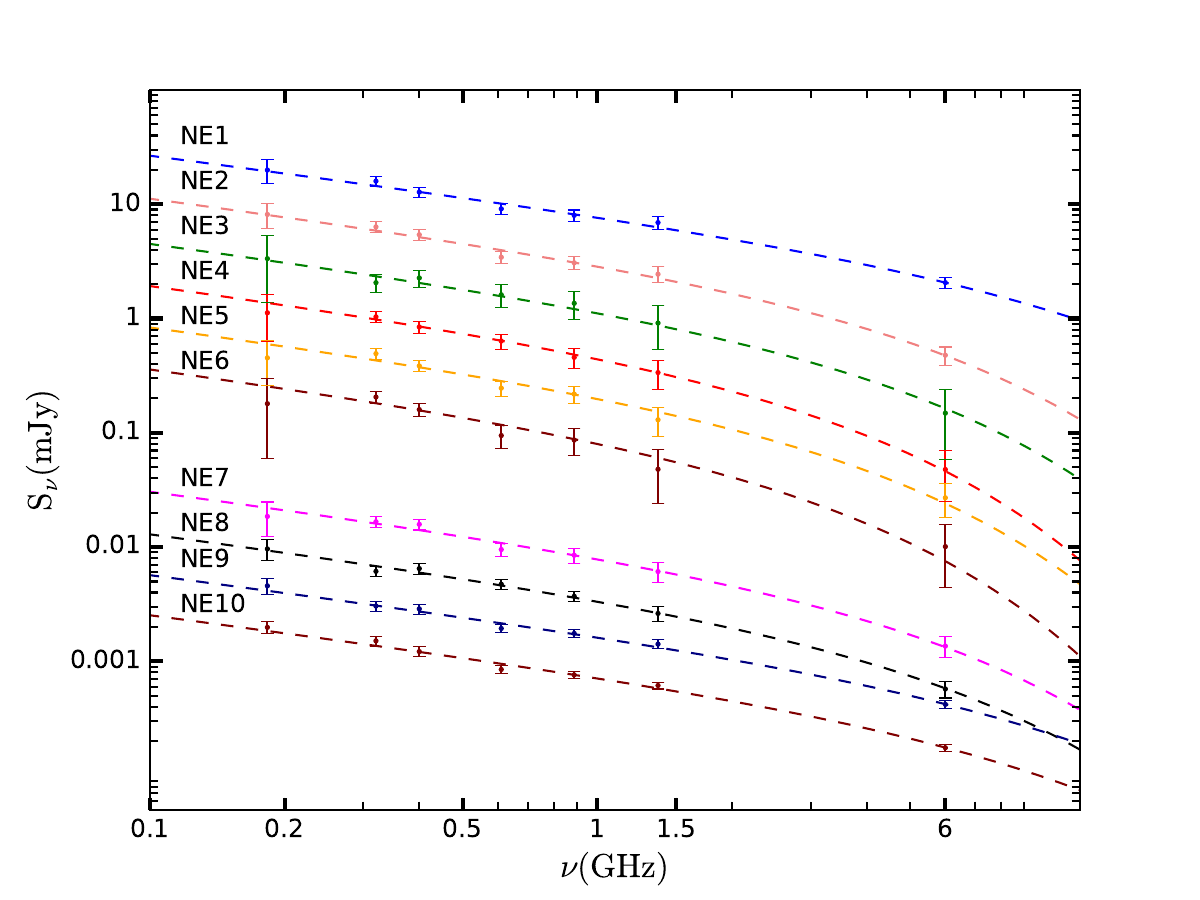}
    \caption{Radio spectra of the different regions from the NE wing (NE1-NE10) fitted with the JP model. The spectra of particular regions are arbitrarily shifted in the ordinate axis to give the appropriate picture of the curvature of the spectra. For details of the flux-density values see Table~\ref{tab3}.}
    \label{fig10}
\end{figure}

\begin{figure}
    \centering
    \includegraphics[width=1\linewidth]{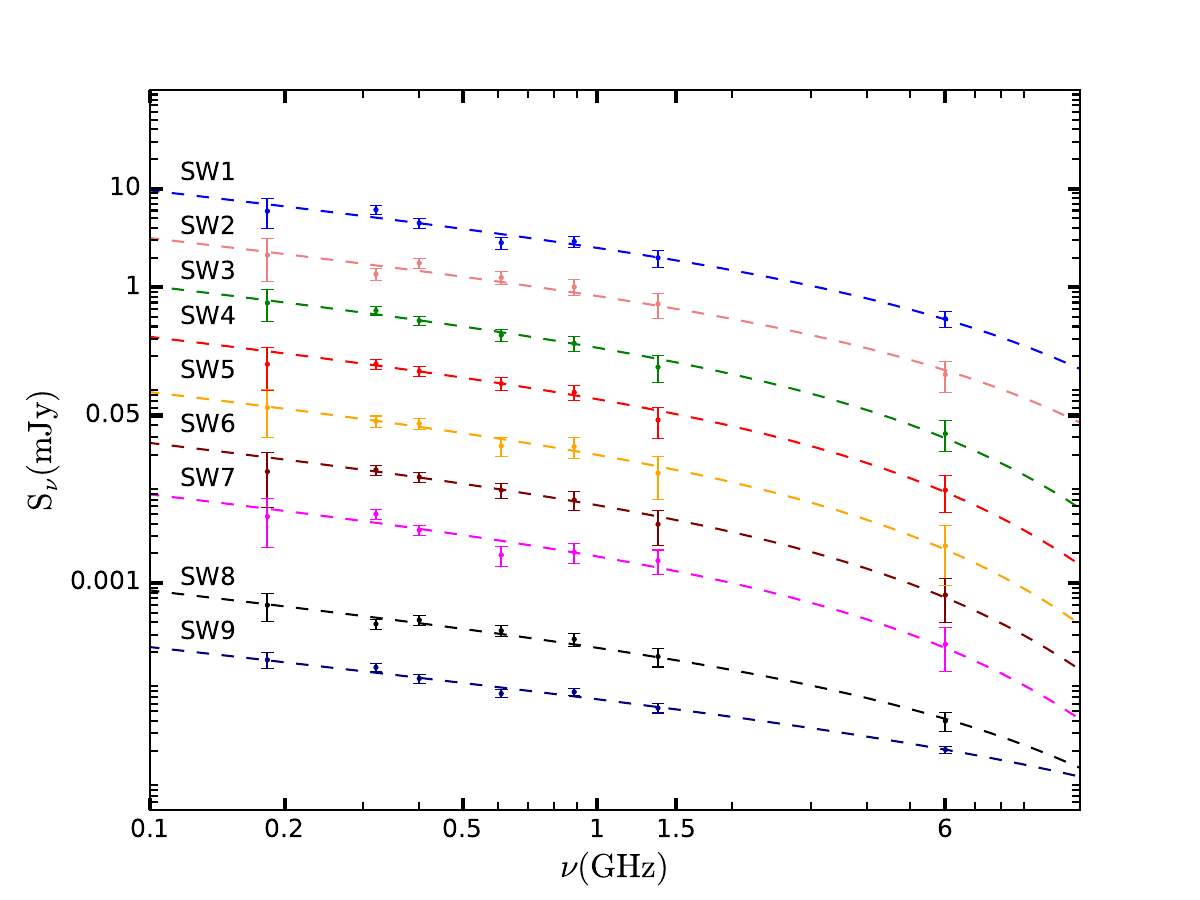}
    \caption{Radio spectra of the different regions from the SW wing (SW1-SW9) fitted with the JP model. The spectra of particular regions are arbitrarily shifted in the ordinate axis to give the appropriate picture of the curvature of the spectra. For details of the flux-density values see Table~\ref {tab3}.}
    \label{fig11}
\end{figure}

 The SYNAGE package (\citealt{1996PhDT........92M}) was used to fit the JP model to the radio spectra. The best-fitting models are shown in Figs~\ref{fig10} and ~\ref{fig11}. The resulting break frequency values of the NE and SW wing regions obtained with the model fit are given in Table~\ref{tab4} and from the pre-fitting results, $\alpha_{\rm{inj}}$ was observed to be close to 0.5, therefore it was fixed at 0.5. The $\nu_{\rm{br}}$ values for both the NE and SW wing regions obtained by the JP model gave comparable or slightly higher $\nu_{\rm{br}}$ values than the highest radio frequency data presented in this article. 

In the radio spectra of the NE wing, as given in Fig.~\ref{fig10}, we see a clear gradient from regions NE1 till NE6, with NE6 showing the steepest spectrum and NE1 showing relatively flatter spectrum. This also indicates the anti-clockwise movement of plasma where the jet moves from region NE6 to region NE1 while sweeping an S-like pattern in the plane of the sky. The regions NE7 to NE10 are much flatter than the regions of NE2--NE6, as also seen from the spectral index map in Fig.~\ref{fig7}. This is most likely the result of these regions being closer to the current direction of southern jet, containing younger plasma. In Fig.~\ref{fig11}, we have the spectra from the regions of the SW wing plotted, and a similar gradient appears as previously seen in the case of the NW wing. In the spectra from SW1 to SW7, the gradient becomes much steeper as we move from regions SW1 to SW7, with SW4--SW7 showing a similar steepening. Here, the spectra of SW8 and SW9 are comparatively flat, which is also consistent with the SI map. This aligns with a northward sweep of the jet in the anticlockwise direction. We therefore observe from the spectra that the jets have undergone reorientation, coming to the current north-south alignment after depositing plasma in the southwestern region, moving through SW7 to SW1, and correspondingly in the northeastern regions, moving through NE6 to NE1. This gives an overall view of the trajectory of the precessing jet, moving in an anticlockwise direction, along the NE and SW wings (see illustration in Fig.~\ref{fig16}).

\begin{figure*}
    \centering
   \includegraphics[width=1\linewidth]{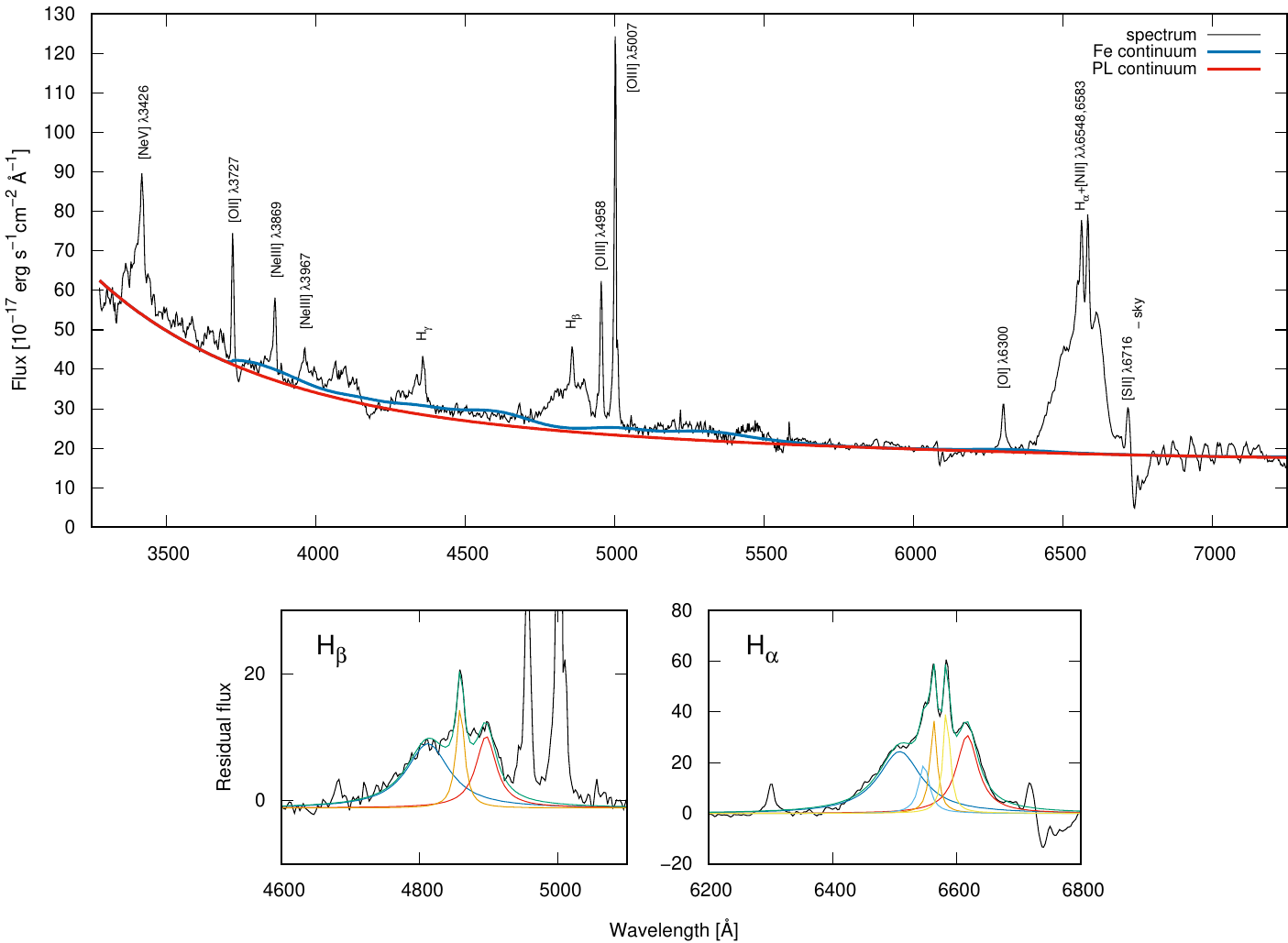}
    \caption{Optical spectrum of PKS\,2300$-$18 obtained from WHT observations. Top panel shows the rest-frame observed spectrum (in black) with power-law continuum (in red) and Fe continuum (in blue). The bottom panels show H$_{\beta}$ and H$_{\alpha}$ regions modelled using a Lorentzian profile. The total fitted flux is shown in green, blue broad components are shown in blue, red broad components are shown in red, and the narrow components are shown in orange. [NII] $\lambda$6548 and [NII] $\lambda$6583 lines in the bottom right panel are shown in light blue and yellow respectively.}
   \label{fig12}
\end{figure*}

\subsubsection{Spectral ageing}
\label{sec3.4.2}
\begin{table}
\begingroup
\caption{Break frequency values and spectral ages of the circular regions computed using the JP particle injection model.}

\label{tab:landscape3}
\setlength{\tabcolsep}{6 pt}
\renewcommand{\arraystretch}{1.6}
\begin{tabular}{cccccccc}
\hline
Region  & Break Freq.  & $\chi^2_{\rm{red}}$  &  $\alpha^*$  & ${B_{\rm{eq}}}$ & Spectral age    \\
 &  (GHz) & & & (\micro G) & (Myr) \\
(1) & (2) & (3) & (4) & (5) & (6) \\
\hline
 SW1  &  $  9.2^{+>100}_{-2.9} $    & 1.13      & 0.88  & $1.06^{+0.20}_{-0.11}$  & $29.00^{+6.04}_{->20.93}$  \\
 SW2  &  $  8.4^{+5.3}_{-5.2}$     &  1.25  &  0.97 &  $1.05^{+0.19}_{-0.11}$  & $30.24^{+21.71}_{-7.55}$  \\
 SW3  & $  5.4^{+>100}_{-1.7}$ & 0.28  & 1.10 & $1.13^{+0.21}_{-0.10}$  & $38.74^{+10.77}_{->30.27}$     \\
 SW4  & $  5.2^{+9.2}_{-2.3} $ & 0.27  &  1.10 & $1.11^{+0.20}_{-0.10}$  & $39.22^{+16.31}_{-16.46}$     \\
 SW5  &  $  5.0^{+>100}_{-2.6} $  & 0. 25     &    1.06 & $0.99^{+0.18}_{-0.09}$  & $38.32^{+20.50}_{->30.27}$   \\
 SW6  &   $ 5.2^{+18.7}_{-2.5} $ & 0.18   &  1.04  & $1.03^{+0.19}_{-0.10}$  & $38.15^{+18.10}_{-21.05}$       \\
 SW7  &   $ 5.3^{+>100}_{-2.0}$  & 1.22   &  1.00 &  $1.05^{+0.20}_{-0.10}$  & $38.07^{+13.22}_{->29.85}$    \\
 SW8  &   $  9.3^{+5.1}_{-5.3} $ &0.65  &  0.91 & $1.18^{+0.23}_{-0.11}$  & $29.97^{+18.36}_{-6.63}$        \\
 SW9  &    $  30.7^{+>100}_{-16.3} $ &1.10  &  1.10 & $1.50^{+0.28}_{-0.14}$  & $17.74^{+9.15}_{->9.37}$     \\
 NE1  &   $ 19.5 ^{+>100}_{-9.4} $    & 0.48  &  0.92 & $1.21^{+0.24}_{-0.12}$  & $20.87^{+9.78}_{->12.74}$     \\
 NE2  &     $  7.7^{+41}_{-2.2} $    & 0.54   &  1.00 & $1.25^{+0.24}_{-0.12}$  & $33.58^{+8.26}_{-21.21}$     \\ 
 NE3  &    $  6.6^{+5.6}_{-3.8} $    & 0.45 &    1.10 & $1.16^{+0.21}_{-0.11}$  & $35.36^{+20.61}_{-10.27}$       \\
 NE4  &    $  4.8^{+12.4}_{-2.1} $    & 0.11   &   1.00 & $1.05^{+0.20}_{-0.10}$  & $40.00^{+16.71}_{-19.66}$      \\
 NE5  &  $  5.5^{+>100}_{-1.6} $    & 0.69   &  1.03 & $1.08^{+0.21}_{-0.10}$  & $37.76^{+9.90}_{->29.45}$       \\
 NE6  &     $  4.3^{+>100}_{-2.1} $    & 0.57  & 0.92 & $0.88^{+0.15}_{-0.10}$  & $39.43^{+19.23}_{->31.82}$          \\
 NE7  &    $  8.0^{+47.1}_{-2.6} $    & 0.53  &   1.00 & $1.16^{+0.22}_{-0.11}$  & $32.12^{+9.22}_{-20.31}$        \\
 NE8  &   $  8.2^{+5.0}_{-3.2} $    & 0.36  &  0.90 & $1.17^{+0.23}_{-0.11}$  & $31.82^{+11.32}_{-7.61}$   \\
 NE9  & $  17.7^{+51.7}_{-7.7} $    & 0.57     &    0.80 & $1.26^{+0.26}_{-0.12}$  & $22.20^{+8.99}_{-11.36}$     \\
 NE10 &    $  15.5^{+33.4}_{-4.7} $    & 0.88     & 0.83 & $1.46^{+0.29}_{-0.14}$  & $24.79^{+6.17}_{-11.23}$  \\ 
 C1 &  $33.1^ {+>100}_{-20.4} $      & 2.09 & 0.62 & $1.82^{+0.39}_{-0.19}$  & $17.81^{+11.53}_{->9.12}$   \\
 C2  &   $28.6^ {+30}_{-16.4}  $   &  1.49   & 0.64 & $2.19^{+0.45}_{-0.23}$  & $19.54^{+10.26}_{-6.02}$ \\
 C3  &     $ 26.6^ {+>100}_{-7.8} $  &   0.92  & 0.73 & $1.94^{+0.39}_{-0.20}$  & $20.05^{+4.02}_{->11.02}$   \\
 C4 &    $ 21.3^ {+>100}_{-6.2} $   &   0.65  & 0.63 & $1.76^{+0.37}_{-0.19}$  & $22.07^{+4.78}_{->13.04}$   \\

\hline
\end{tabular}
\begin{tablenotes}\footnotesize
\item *The alpha ($\alpha$), i.e. the mean spectral index values of the individual regions were estimated from the SI map in Fig.\ref{fig7}.
\end{tablenotes}
\label{tab4}
\endgroup
\end{table}

We estimate here the magnetic field strength and synchrotron age of the NE and SW wing regions from NE1-NE10 and SW1-SW9, respectively. The spectral age is calculated using \citet{1980ARA&A..18..165M} as follows: 

\begin{equation}
 \tau_{\rm{rad}}=1590 \frac{B_{\rm{eq}}^{0.5}}{B_{\rm{eq}}^2 + B_{\rm{CMB}}^2} (\nu_{\rm{br}}(1+z))^{-0.5} {\mathrm{Myr}}
 \end{equation}

{$B_{\rm{CMB}} = 3.18(1 + z)^2$ is the magnetic field strength analogous to the cosmic microwave background radiation. The magnetic field strength of the wings, i.e., ${B_{\rm{eq}}}$, and $B_{\rm{CMB}}$ are expressed in units of $\micro$G. The spectral break frequency $\nu_{\rm{br}}$ (in GHz) is above which the radio spectrum steepens from the initial power-law spectrum, given by $\alpha_{\rm{inj}}$=($\gamma-1$)/2. 
The magnetic field calculations (${B_{\rm{eq}}}$) were done in accordance with the revised equipartition arguments provided by \citet{2005AN....326..414B} (see equation 3 therein). Magnetic field strength can be estimated directly for a source if it is simultaneously detected at radio and X-ray frequencies. Although our radio source was observed with ROSAT, its angular resolution is insufficient to distinguish the emission originating from the wings from the strong X-ray emission generated by the radio core. 

\begin{figure*}
    \centering
    \includegraphics[width=1\linewidth]{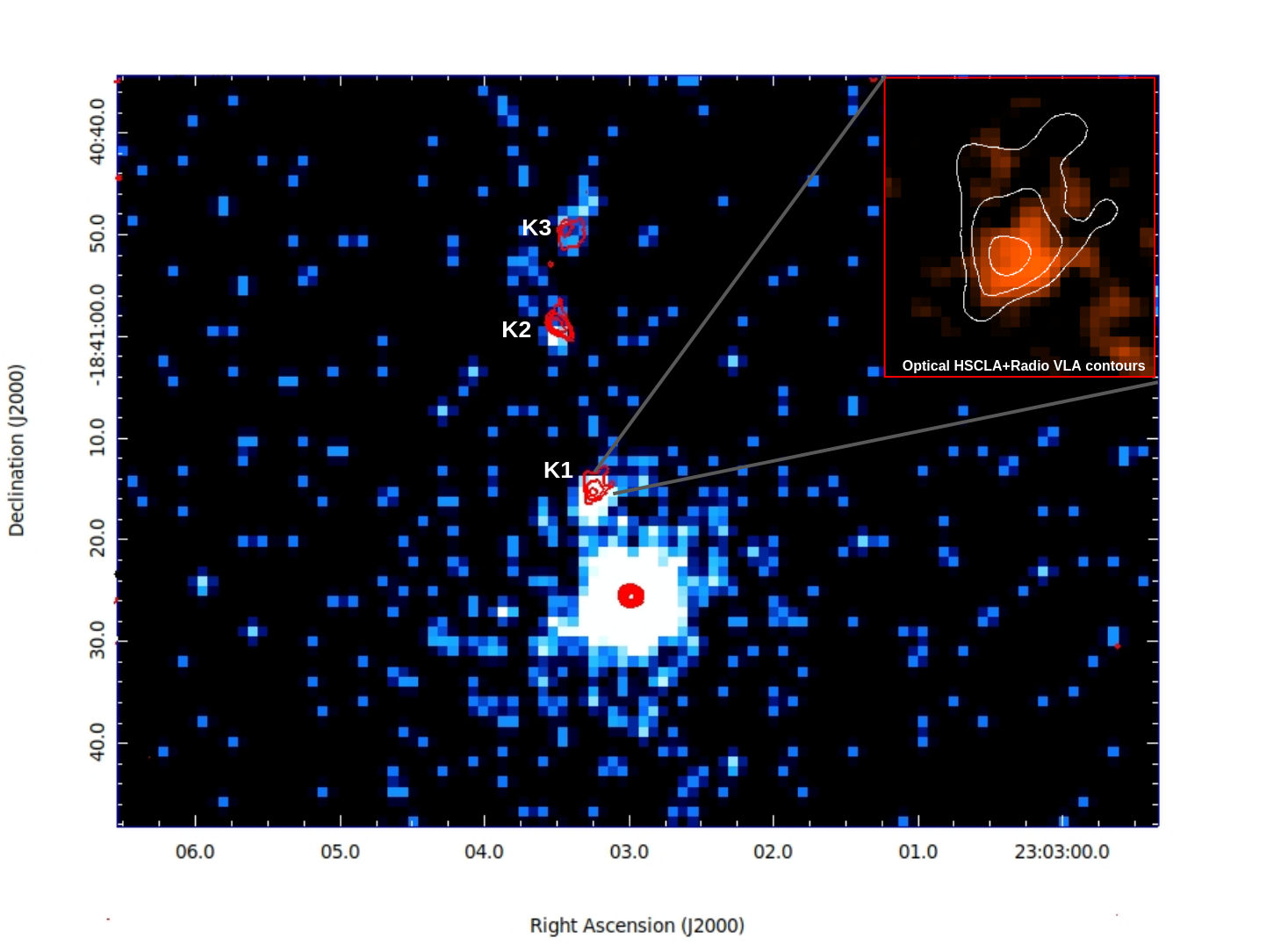}
    \caption{X-ray image from Chandra of the host PKS\,2300$-$18 as seen in blue. The red contours overlaid on top of the X-ray map are VLA contours at 5 GHz with 0.7 arcsec resolution. The radio contours show the location of the core and three knots of the northern jet. The radio knots coincide with points of enhanced X-ray emission. In the inset towards the top-right corner we have the first radio knot, K1, as seen in the HSCLA -r band image in orange, overlaid with radio VLA 5 GHz contours.}
    \label{fig14}
\end{figure*}

For the calculation, the following assumptions were made: (i) the ratio of proton-to-electron number densities K = 0 (a relativistic electron-positron plasma), (ii) the polarised emission originates from regular magnetic field with all possible inclinations, and (iii) degree of polarisation is 7 per cent (for detail see Sect. 3.1.2). We made use of the BFIELD program developed by \citet{2005AN....326..414B}  to apply their revised calculations and obtained the magnetic field strength of the 23 different regions within the radio structure as marked in Fig.~\ref{fig8}. The volume of the circular regions in the wings was estimated using a cylindrical geometry. The calculated values of magnetic field strength and spectral age of the respective regions are given in Tab.~\ref{tab4}. It is also observed that the magnetic field in the wing regions are mostly uniform, ranging between 0.88–2.19 µG and the median value is $\sim$2.1 times lower than the minimum magnetic field value of $B_{\rm{CMB}}$/$\sqrt{3}$ = 2.34 \micro G, that minimises the radiative losses and maximises the lifetime of the source providing an upper limit. In the case of the SW wing, the spectral age from SW1--SW7 range between 29--39 Myr, with the youngest plasma being in the region SW1, which is closest to the current location of the northern jet, with SW7 containing comparitively older plasma. We see a similar gradient in the ageing of the NW wing between NE1 and NE6, ranging from 21--40 Myr, with NE6 containing the oldest plasma of 39 Myr and NE1 containing the youngest plasma of 21 Myr. NE6 is farthest from the southern jet, whereas NE1 is closest to the southern jet, so this trend in spectral age confirms the anticlockwise movement of plasma in the wings. The spectral ages of C1, C2, C3, and C4, located in the northern and the southern jet regions have spectral ages between 17--22 Myr and magnetic field in the range of 1.8--2.2 $\micro$G. The magnetic field is higher in these regions compared to the NE and SW wings and the spectral ages are lower, given that it is along the current North-South orientation of the jets. Nonetheless, these ages might still be overestimated because of the presence of diffuse background plasma originating from earlier emission(s) due to the former orientation of the jets during precession.

We hence observe that the oldest plasma in both the NE and SW wings are of similar age, supporting our assessment of jet reorientation taking place in the source. Based on our analysis, we interpret that the NE wing was formed as the southern jet gradually moved from its earliest position at NE6 to its current position at NE1. Similarly, the northern jet progressed from SW7 to SW1, resulting in the formation of the SW wing.

\section{Optical properties of the central AGN}
\label{sec4}

The rest-frame optical spectrum shown in Fig.~\ref{fig12} clearly exhibits a quasar spectrum, characterised by a power-law continuum and a combination of broad and narrow emission lines. In particular, the H${\alpha}$ and H${\beta}$ lines feature broad double-peaked components.  
The power law (PL) and the iron continuum (Fe) were fitted with PyQSOFit \citep{PyQSOFit2018}. The emission lines measured in the residual spectrum presented extended wings and therefore were fitted with a Lorentzian profile. 
The H$_{\alpha}$ and H$_{\beta}$ lines were fitted with a combination of one narrow and two broad components. The H$_{\alpha}$ line was fitted simultaneously with [NII]$\lambda\lambda$6548,6583 and [OI]$\lambda$6300 narrow lines that were blended with H$_{\alpha}$. [SII]$\lambda\lambda$6717,6731 lines were affected by the telluric sky line and were not taken into account. The H$_{\beta}$ line was fitted with [OIII]$\lambda\lambda$4959,5007 lines and HeII $\lambda$4687. [OIII]$\lambda$5007 was affected by the bright sky line and therefore was not taken into account in further analysis. 
The fits obtained for both Balmer lines were consistent and they were used to obtain information about the central AGN. The fluxes, widths, and the shifts of the broad components with respect to the narrow ones are presented in Table~\ref{tab5}.
The optical quasar outshines the host galaxy; therefore, it was not possible to extract the host galaxy's contribution from the optical spectrum.

\begin{table}
\caption{The H$_{\alpha}$ and H$_{\beta}$ parameters of the rest-frame optical emission lines of  PKS\,2300$-$18.}
\centering
\begin{tabular}{ccccc}
\hline
Line & Component & Flux & FWHM & Shift $v_d$ \\ 
& & (10$^{-15}$ erg s$^{-1}$ cm$^{-2}$) & (km s$^{-1}$) & (km s$^{-1}$) \\ 
\hline
 & broad blue & 13.19$\pm$0.84 & 5490$\pm$1300 & -2840$\pm$96 \\ 
H$_{\beta}$ & narrow & 3.44$\pm$0.23 & 820$\pm$110 & \\ 
 & broad red & 6.28$\pm$0.08 & 2130$\pm$540 & 2331$\pm$8 \\ 
 & total broad & 19.47$\pm$0.92 & 9020$\pm$1890 & \\
\hline
 & broad blue & 32.8$\pm$3.0 & 3970$\pm$400 & -2520$\pm$250 \\ 
H$_{\alpha}$ & narrow & 7.0$\pm$1.3 & 470$\pm$280 & \\ 
 & broad red & 21.78$\pm$1.5 & 2000$\pm$280 & 2457$\pm$46 \\ 
 & total broad & 54.6$\pm$4.5 & 8015$\pm$950 & \\
\hline
\end{tabular}

\label{tab5}

\end{table}

Monochromatic luminosity at 5100\AA\ is often used to estimate the virial mass of the SMBH, size of the broad-line region (BLR), the AGN bolometric luminosity and Eddington ratio. 
The measured flux density at 5100\AA\ = (22.83$\pm$0.55) $\times 10^{-17}$ erg s$^{-1}$ cm$^{-2}$ \AA$^{-1}$ and the source luminosity distance, $D_{L}$ = 601.7\ Mpc gives the monochromatic luminosity $\lambda L_{\lambda}$(5100) = (5.04$\pm$0.13) $\times 10^{43}$ erg s$^{-1}$. 
Using the virial mass-luminosity relation of \citet[][equation 5 therein]{Greene&Ho2005} and the width of the broad component of H$_{\beta}$ = 9020 km s$^{-1}$, derived according to the procedure described by \citet{PetersonEtAl2004}, the total black hole (BH) mass in PKS\,2300$-$18 is = (2.31$\pm$1.02)$\times 10^8$ M$_{\odot}$. The size of the BLR, estimated based on the \citet{Greene&Ho2005} relation, is 13.9$\pm$0.2 light-days.  

Total bolometric luminosity can be calculated assuming a simple relation between monochromatic luminosity $\lambda$L$_{\lambda}$(5100), and the AGN bolometric luminosity in the form of L$_{bol}=C_{\lambda}\lambda L_{\lambda}$. 
With the value of C$_{5100}$=9 adopted after \citet{KaspiEtAl2000}, the bolometric luminosity of PKS\,2300$-$18 = (4.54$\pm$0.1) $\times 10^{44}$ erg s$^{-1}$, and the Eddington ratio is 0.015$\pm$0.007.

Double-peaked broad emission lines are a rare feature among AGNs, and radio sources in particular. They are often explained by the presence of a binary black hole \citep[BBH: e.g.][]{Gaskell1983,Popovic2012}, irradiated disc \citep[]{EracleousHalpern1994}, or irradiated disc with a spiral perturbation \citep[]{ChakrabartiWiita1994,GilbertEtAl1999}. With only one-epoch spectral data, these scenarios cannot be easily distinguished. 
However, in the case of PKS\,2300$-$18, where precession of the jets and radio and optical variability are observed, the assumption about the presence of a BBH system seems plausible.

There may be different configurations of the BBH system, with (1) both BH active and having their own BLRs, (2) both BH active but with a common BLR, and (3) only one BH active (see, e.g. \citealt{Popovic2012}). With the available data, the first scenario could be tested.

As shown by \citet{Greene&Ho2005}, the $\lambda$L$_{\lambda}$(5100) correlates well with the  H$_{\alpha}$ and H$_{\beta}$ luminosities. Assuming that the double-peaked broad emission lines result from the presence of a BBH with separate BLRs, each broad component can be interpreted as being emitted by a different AGN. In that case, the ratio of the luminosities of broad line components should be the same as the ratio of $\lambda$L$_{\lambda}$(5100) of each AGN. BLRs radii would be then 15.2$\pm$1.1 and 9.5$\pm$0.1 light-days based on H$_{\beta}$ (14.1$\pm$1.3 and 10.8$\pm$0.7 light-days based on H$_{\alpha}$). 
From the separation of broad components, the separation of the BHs in the system can be found as follows (\citealt{2024MNRAS.530.4902S}): 

\begin{equation}
 {d_{BBH}}\leq \frac{G*(M_{BH1}+M_{BH2})}{\Delta v_d^2}  ,
\end{equation}

This separation from H$_{\beta}$ measurements would be equal to 17.4 light-days and 10.1 light-days according to H$_{\alpha}$ data. These values are smaller than the sum of the BLRs radii, and thus BHs would be surrounded by a common BLR. With these simple estimates the scenario about separate BLRs contributing to the double-peaked broad emission lines in PKS\,2300$-$18 can be ruled out. However, the remaining two other possibilities cannot be easily distinguished as  PKS\,2300$-$18 is a variable source (see Sections~\ref{sec1} and ~\ref{sec3.3}) and many quasars also show variations in broad emission lines; therefore, estimates based on one-epoch observations can be biased. Visual comparison with previously published data from \citet{1984MNRAS.207...55H} and \citet{EracleousHalpern1994} may suggest spectral variability, and future systematic spectral observations and reverberation mapping can help to clarify the scenario of the structure of the line-emitting gas and help to distinguish between a single-irradiated disc and the SMBH binary scenario \citep[e.g.][]{ShenLoeb2010}.

\begin{figure}
  \centering
    \includegraphics[width=0.77\linewidth]{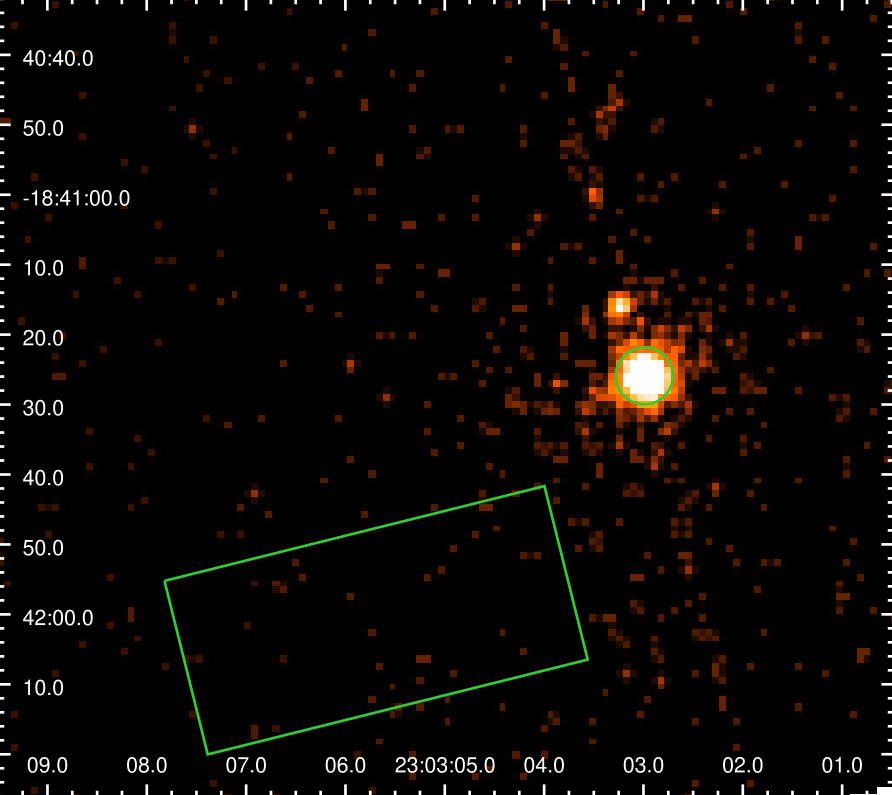}
    \includegraphics[width=0.85\linewidth]{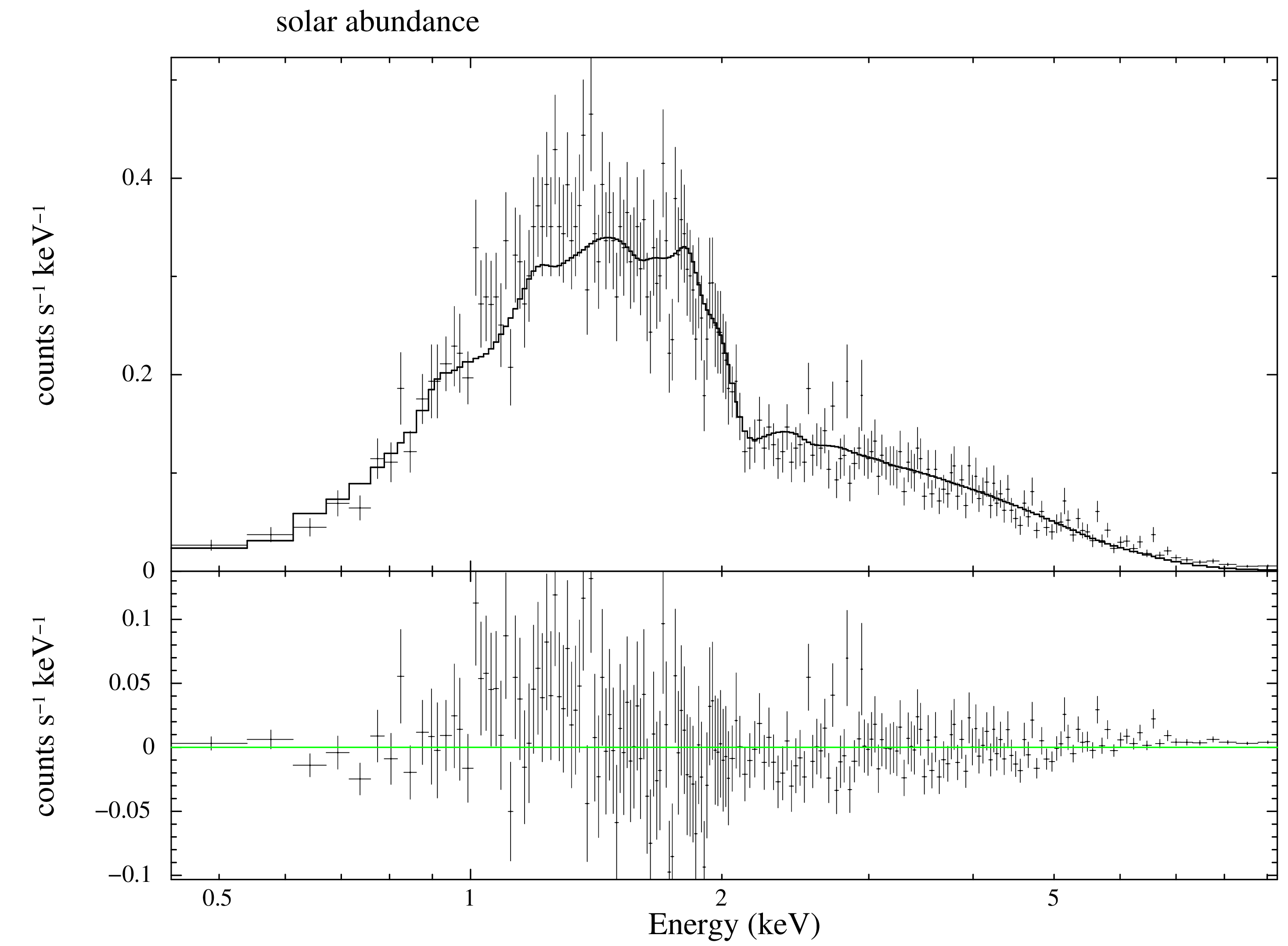}
    \includegraphics[width=0.85\linewidth]{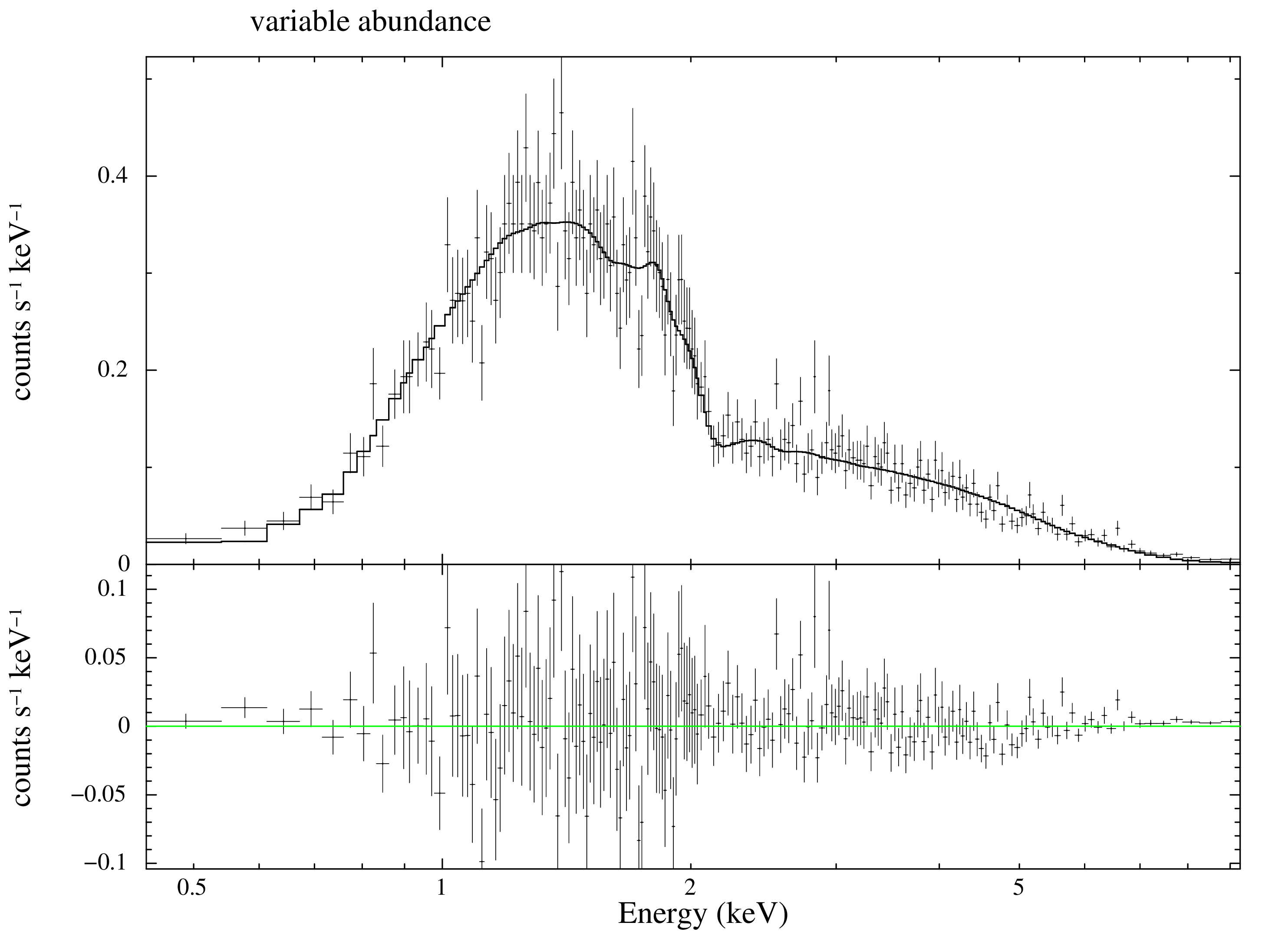}
    \caption{Top: regions used in the spectral analysis of the core of PKS\,2300$-$18 from Chandra X-ray data. The core region is marked with a green circle and the background region with a green rectangle. Bottom: extracted Chandra X-ray spectrum of the core region with the fitted solar abundance and variable abundance model and residuals. For detailed discussion see Section~\ref{sec5}.}
    \label{fig13}
\end{figure}

\section{X-ray study}
\label{sec5}

In the X-ray broadband (0.2$-$7\,keV) Chandra image of PKS\,2300-18, the radio core coincides with a bright point-like source, the extent of which reaches 8 arcsec, which is significantly larger than the PSF of the Advanced CCD Imaging Spectrometer (ACIS) instrument (Fig.~\ref{fig13}). This suggests that the X-ray image captures not only the central AGN but also the hot interstellar medium of the host galaxy. Toward the north, a few more compact X-ray sources are visible that coincide well with the radio knots K1-K3 in the Chandra image as seen in  Fig.~\ref{fig14} . The knot K1 is also visible in the optical HSCLA r-band image, as seen in the insert in Fig.~\ref{fig14}, making it the first multiwavelength detection of a knot in an S-shaped RG jet. The radio to optical emission (and X-rays in some cases) can be attributed to synchrotron radiation from the jet extended at higher frequencies, produced by relativistic electrons interacting with the jet’s magnetic field. However, such X-ray emission in the knots can also stem from the inverse Compton scattering of CMB photons, while the optical emission might originate from upscattered CMB photons when they interact with the relativistic electrons in the jet. An interesting observation here is that a visible lag can be seen between the radio contour and the X-ray peak, seen most clearly in knot K1 and to a lesser extent in knots K2 and K3.

We extracted spectra of all four sources (knots and centre), but only in the case of the host galaxy (or core region) was a more detailed spectral analysis possible. For this spectrum, we used a model consisting of a thermal and a power-law component accounting for the hot gas in the host galaxy/core region and the central source, respectively. The thermal component is described by the {\it mekal} model \citep{mewe85,kaastra92} of thermal bremsstrahlung with additional emission lines. The metallicity in this model was set to solar. The absorption in our Galaxy was taken into account via the {\it wabs} model with a fixed value of column density towards the source.

The derived model describes the spectrum of the core region well, except for the energy range between 1 and 1.5\,keV, in which a significant excess of X-ray emission over the model can be observed. This could suggest that an additional thermal component of a very high temperature (describing the hot gas in the core region) is needed. However, such a modification of the model did not improve the fit and the new parameter could not be constrained. Much better results were obtained by allowing the metallicity of the mekal model to vary. In this case, however, a good model fit to the spectrum required metallicity to be as low as around 3 per cent solar. Such a level of metal abundance certainly needs explanation. We note here that the radius of our spectral region of 4 arcsec corresponds to around 9\,kpc, which means that we probe the conditions in the direct vicinity of the AGN. Recently, \citet{2024MNRAS.528.1863R} found similar low metallicity at the core of a low-redshift quasar (within 10\,kpc), which they explained as a potential result of photo-ionization of the hot atmosphere by the quasar emission. We argue that this might also be the case for PKS2300-18. In Table~\ref{xmodel} and Fig.~\ref{fig13} we present the parameters of the fitted models (both variable and solar metallicity) and their plots with residuals, respectively.

\begin{table}
    \caption{Parameters of the model fit to the spectrum of the core region.}
    \label{xmodel}
    \centering
    \begin{tabular}{ccc}
        \hline\hline
         & low Z & solar Z\\
         \hline
        kT [keV] & 0.61$^{+0.22}_{-0.16}$ & 0.32$^{+0.28}_{-0.04}$\\
        S$_{\rm kT}$ [10$^{-12}${\rm erg}\,{\rm cm}$^{-2}$s$^{-1}$] & 4.83$^{+5.85}_{-2.49}$ & 2.73$^{+2.19}_{-1.15}$\\
        $\Gamma$ & 1.04$^{+0.15}_{-0.19}$ & 1.41$\pm$0.05 \\
        S$_{\rm \Gamma}$ [10$^{-12}${\rm erg}\,{\rm cm}$^{-2}$s$^{-1}$] & 14.4$^{+11.3}_{-6.3}$ & 14.7$^{+2.3}_{-2.0}$\\
        red. $\chi^2$ & 1.096  & 1.328 \\
        d.o.f. & 193 & 194 \\
        \hline
    \end{tabular}
    \begin{tablenotes}\footnotesize
   \item In our model fits we used the weighted average value of $N_{\rm H}$=1.88$\cdot$10$^{20}$\,cm$^{-2}$ after the HI4PI survey; \citealt{2016A&A...594A.116H}. The fluxes are derived in the range of 0.3-10\,keV.
     \end{tablenotes}
\end{table}

\section{Discussion}
\label{sec6}

\subsection{Formation of S-shaped jet morphology}
\label{sec6.1}

The characteristic curvature of radio jets in the case of winged radio galaxies is observed in the form of X-, S-, and Z-shaped morphologies (\citealt{2017MNRAS.465.4772R}; \citealt{2020MNRAS.495.1271C}; \citealt{2022MNRAS.512.4308B};  \citealt{2023MNRAS.523.1648M}; \citealt{2024ApJ...969..156S}). In most of these sources, one pair of lobes undergoes active evolution, often leading to hotspot formation while the accompanying pair of wings or the secondary pair of lobes have low surface brightness emission and are relatively diffuse and extended. Until recently, distinctively recognizing winged morphologies in low-resolution observations was challenging; however, with the advent of high-sensitivity and deep radio surveys, it has become easier to distinguish between different classes of winged RGs. X- shaped sources are typically formed as a result of backflowing jet plasma deflected laterally by an asymmetric medium (\citealt{1984MNRAS.210..929L}; \citealt{2002A&A...394...39C}; \citealt{2020MNRAS.495.1271C}; \citealt{2024FrASS..1171101G}) or by the sudden spin flip of an SMBH post-merger (\citealt{2002Sci...297.1310M}; \citealt{2002PhDT.......178R}), whereas the Z- and S-shaped radio sources are often considered to be the result of underlying jet precession (\citealt{1982ApJ...262..478G}; \citealt{2002PhDT.......178R}; \citealt{2020MNRAS.499.5765H}; \citealt{2023MNRAS.521.2593H}). As marked by \citet{2019MNRAS.482..240K}, the best indicators of jet precession are (i) a jet at the edge of the lobe, (ii) curvature in the jet path, (iii) S-shaped symmetry, and (iv) multiple or complex hotspots. These indicators make the S-shaped morphology one of the most suitable classes among winged RGs for studying jet precession.\\

When studying such S-shaped sources, it is important to be cautious of the environmental factors that might lead to an S-shaped morphology other than the conventional jet precession. Sources displaying helical jets have been found in X-ray binaries (\citealt{1981Natur.290..100H}), planetary nebulae (\citealt{1993ApJ...415L.135L}) and AGN jets (\citealt{2018MNRAS.480.3644R}; \citealt{2024A&A...688A..86U}).
A prototypical helical jet can be described by three different jet structures as elucidated by \citet{1997VA.....41...71S}. The first model consists of a ballistic helical jet, where the individual jet fluid elements follow a straight path but the overall structure becomes helical as the direction of ejection of the fluid elements changes periodically; these jets demonstrate precession.  In the second model, the helical jets are bent as a whole and the fluid elements follow a common twisted trajectory, delineated by a curved jet axis. Jet bending of this type can be seen when the jets undergo intrusion due to collision with gas clouds or interaction with transverse winds or a dense ambient medium. In RGs like 3C\,433 and 4C\,65.15, which exhibit jets with a hybrid FR morphology, Chandra observations reveal that the FRI structure observed on one side of these hybrid sources arises from a powerful jet interacting with a comparatively dense medium (\citealt{1983AJ.....88...40V}; \citealt{2009ApJ...695..755M}). Also such jet bending can be observed in the case of 3C 321 (\citealt{2008ApJ...675.1057E}), where one of its jets bends after interacting with a neighboring companion galaxy that is undergoing merger with the host galaxy. In the third model, the jets internally have a helical structure.  Such jets are straight as a whole but have fluid elements flowing along helical trajectories within the jet. This type of jet might be produced as a result of Kelvin-Helmholtz instabilities (\citealt{1987ApJ...318...78H}; \citealt{1992ApJ...400L...9H}; \citealt{2006A&A...456..493P}; \citealt{2012A&A...545A..65P}). These instabilities can be observed in transverse oscillations in the M87 jet during VLBA observations, where a shift of the transverse position of the jet on a quasi-periodic 10-yr time-scale was seen consistent with the Kelvin–Helmholtz instability (\citealt{2018ApJ...855..128W}; \citealt{2023MNRAS.526.5949N}). The VLBI observations of  M87 jet also showed a 1-yr period wiggles in multi-epoch  observations (\citealt{2023Galax..11...33R}).

We first consider the case where the S-shaped jets of PKS\,2300$-$18 are bent as a result of local environment interaction and discuss other scenarios thereafter. The S-shaped jets, which are currently orientated in the north-south direction, show inversion-symmetric jet bending. It is statistically unlikely that the northern and southern jet encounter and are obstructed by gas clouds within the confines of the host galaxy located on an axis of about 180 degrees. Furthermore, such obstruction by clouds that are aligned in opposite directions on either side of the galaxy will likely cause jet bending/curvature much closer to the periphery of the host galaxy, however, we only see strong jet curvature at a scale of more than 150 kpc. The formation of multiple jet knots (see Fig.~\ref{fig3}), much further than the confines of the host galaxy, in the case of the northen jet also hints that the jet is not getting obstructed by gas clouds, which would lead to a jet with a diffuse structure rather than a collimated structure as seen in the case of the northern jet. The presence of diffuse emission around the S-shaped jets also indicates that the jets were previously oriented in the northeast and southwest direction and have gone through jet reorientation. Therefore, given the above arguments, the helical jets of our target seem unlikely to be a result of collision with gas clouds.  \\

We now consider the possibility that the observed bending is a result of the pressure gradient within the hot gas halo surrounding the host galaxy. In such a case, the halo should lead to a plane-symmetric effect for both jets (e.g. 3C 98) and an uneven curvature of the jets, rather than the S-shaped inversion symmetry seen in the case of PKS\,2300$-$18. Another potential explanation could be that the Kelvin-Helmholtz instability in the jets is responsible for the observed bending. This instability would most likely give rise to 'wiggles' or sharp small bents in the jets. However, we observe a well-collimated curved jet, which is visible in the case of the northern side of the jet and completely absent on the southern side of the jet. These jets also appear to be curved and not straight as a whole. Therefore a Kelvin-Helmholtz instability seems to be an unlikely explanation for the S-shaped jets seen in our target. The most plausible scenario here remains of a ballistic helical jet that arises due to jet precession. In the high-resolution VLA L-band map, as seen in Fig.~\ref{fig3}, the three jet knots that are visible in the northern jet are seen moving along curved trajectories, which is strongly indicative of plasmoids being ejected in different directions. Furthermore, there is low surface brightness extended emission along the north-eastern and south-western side of the S-shaped jets that is caused by jet reorientation. Such large-scale jet reorientation can only be explained by a jet precession.

\subsection{Kinematical modelling of jet precession}
\label{sec6.2}

In this section, we proceed towards modelling the precession of the radio jet in the case of PKS\,2300$-$18. We modelled the jet using the kinematic jet precession model of \citet[][]{1982ApJ...262..478G}, originally developed by \citet[][]{1981ApJ...246L.141H} to model the jets of the SS433 microquasar. In this model, symmetric jets are launched along an axis that traces a cone throughout a precession period P. The model parameters primarily include the following: the precession cone half opening angle ($\psi$), the cone inclination angle (i), the position angle ($\theta$), and the jet advance speed ($\beta$). The proper motion plots were derived from the visual inspection, by varying the values of the model parameters to match the radio map of PKS\,2300$-$18 at JVLA 6 GHz given in Fig.~\ref{fig15}. The best-fitting model values, which were fitted to multiple high-resolution maps including the VLA-L band A and C configuration map with the highest resolution for constraining the degeneracy, are $i = 67\degree$, $\psi = 66\degree$, $\theta = 119\degree $ ; note that these values are similar to the ones obtained by \citet{1984MNRAS.207...55H}. To determine the errors associated with our precession model we estimated the range of parameters that can fit the radio morphology by fixing all parameters except one, and then subsequently varying it until the visual fit became noticeably worse (\citealt{2008MNRAS.388.1457S}), this step was then followed for the rest of the parameters. The parameter ranges obtained using this method are $i = 67\degree\pm4\degree$, $\psi = 66\degree\pm4\degree$, $\theta = 119\degree \pm 4\degree $ . This gave a best fit for P in the range $12\pm8$ Myr. The $\beta$ here was confined to be below 0.2c, as values higher than that lead to asymmetric lobes, making one side lobe much larger than the other. The lower limit was set as 0.01c. The precession period obtained here is a few Myr lower than the maximum spectral age obtained for the source, which might hint that the jets have gone through more than one precession cycle. The upper limit of $\beta$ is much lower than the ones obtained from the analysis of the VLBA pc-scale jets in Section ~\ref{sec3.2}. This may be attributed to changes in the large-scale precessing jets, which have been observed to decelerate as they interact with the intergalactic medium in simulation studies (\citealt{2022MNRAS.514.5625G}). As a result, while the pc-scale precessing jets might still have relativistic speeds, their bulk motion at the kpc scale can remain mildly relativistic.

\begin{figure}
    \centering
    \includegraphics[width=1\linewidth]{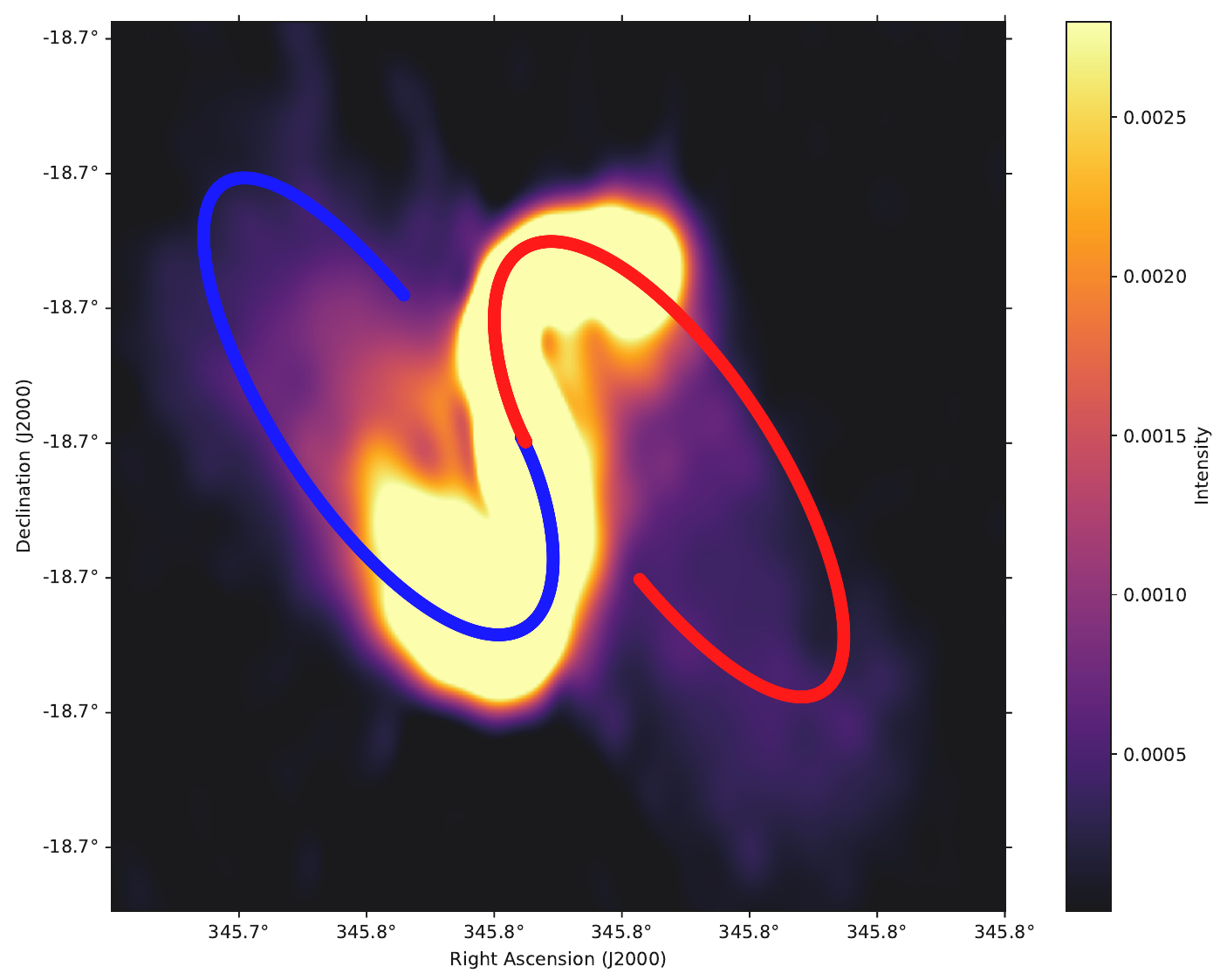}
    \caption{Kinematic jet precession model overlaid on top of the JVLA 6-GHz map. The colour gradient represents flux density in the units of Jy \( \text{beam}^{-1} \). The red and blue lines represent jet and counter jet respectively.}
    \label{fig15}
\end{figure}

\begin{figure*}
    \centering
    \includegraphics[width=1\linewidth]{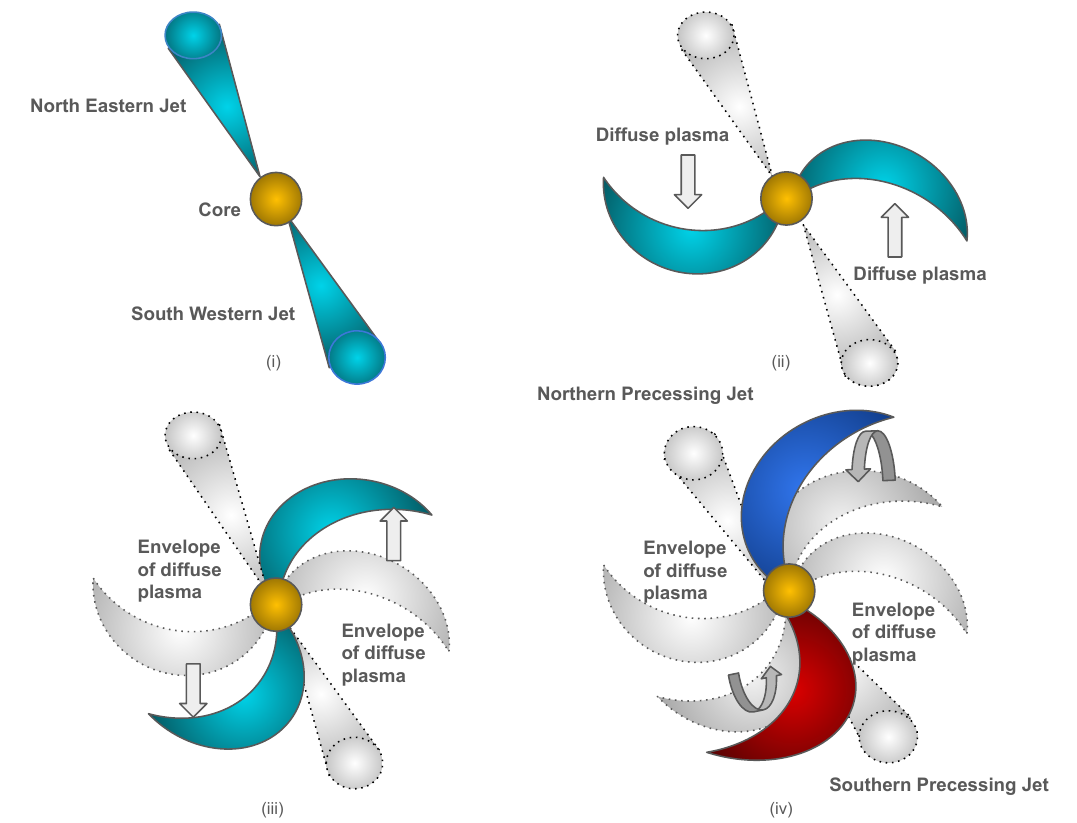}
    \caption{Precessing jet model illustration showing the movement of the jets in PKS\,2300$-$18. Panels (ii) and (iii) encapsulate the anti-clockwise movement of the precessing jets, and panel (iv) following the anti-clockwise motion showcases the motion of the northern precessing jet towards the line of sight (blue), and southern precessing jet away from the line of sight (red).}
    \label{fig16}
\end{figure*}

\subsection{Mechanisms for jet precession}
\label{sec6.3}
Jet precession resulting in S-shaped radio jets is likely to occur in a binary SMBH system (\citealt{1980Natur.287..307B}).
Within such a binary system, the spin axes of the BHs are likely to be misaligned, unless they have both accreted a significant amount of gas with fixed angular momentum. The spin axes will then subsequently undergo geodetic precession and for the more massive BH the precession period $P_{\text{prec}}$ can be given by the following:

\begin{equation}
P_{\text{prec}} \sim 600 r_{16}^{5/2} \left( \frac{M}{m} \right) {M_8}^{-3/2} 
    yr
\end{equation}

where $r_{16}$ is the separation in units of $10^{16}$ cm, $M_{8}$ is the mass of the primary SMBH in units of $10^8 M_\odot$ and (M/m) is the mass ratio of the primary to the secondary SMBH (\citealt{1980Natur.287..307B}). Considering a scenario where BH mass ratios lie in between 1 and 10, primary BH mass of $2.31\times 10^8$ M$_{\odot}$ (from Section~\ref{sec4}), and the precession period of 12 Myr from the kinematical model in Section ~\ref{sec6.2}, we obtain binary separation to be $\sim 335$ and $\sim 133$ light days, respectively. This indicates the possibility that the binary has shrunk to a sub-pc scale separation, due to which even in the VLBI observations (as seen in Fig.~\ref{fig5}) it would be impossible to separate two different components at the centre. The orbital period of such a BBH, assuming a Keplerian orbit, can be given by the following:

\begin{equation}
P_{\text{orb}} \sim 1.6r_{16}^{3/2}{M_8}^{-1/2} 
    yr
\end{equation}

This results in an orbital period of $\sim 856$ and $\sim 212$ years for binary SMBH mass ratios of 1 and 10 respectively, and can be categorised as a wide binary system. Besides a BBH system, a single AGN model can also explain the helical motion of BH jets. Based on the \citet{1980ApJ...238L.129S} model for SS 433, a tilted accretion disc can also cause jet precession. According to this model, if the accretion disc is tilted with respect to the equator of the central SMBH, i.e. the rotation axis does not align with the central BH spin, then the Lens-Thirring effect (\citealt{LenseandThirring}) can cause the disc material inside a specific inner radius to align with the BH's equatorial plane. Meanwhile, the outer part of the disc, which possesses significant angular momentum, will preserve its orientation and induce precession of both the central BH and the inner disc (\citealt{1975ApJ...195L..65B}). Some material of the inner disc is assumed to be ejected out along the BH's spin axis, forming a precessing jet. This model of a tilted accretion disc around a single AGN also gives a relation between the precession period and optical luminosity as given by \citet{1990A&A...229..424L}.  \\

We have used this period-luminosity relation, where the absolute B-band magnitude of our target is $M_{abs}$ =--20.51, which gives an estimated precession period in the range of  $\sim$ 6--20 Myr. This does agree well with the dynamical age from the jet precession model, but differs by a few Myr from the maximum spectral age of 40 Myr. It also remains a possibility that the accretion disc, in the case of  PKS\,2300$-$18, becomes wrapped or tilted due to nonuniform irradiation from the AGN which can lead to jet precession (\citealt{1997MNRAS.292..136P}). As we have not resolved two separate optical cores and VLBI observations are limited by their resolution, it is hard to rule out one of the above precession scenarios. Therefore, jet precession in PKS\,2300$-$18 might be caused either by SMBH binary or by a wrapped/tilted accretion disc.

\subsection{ Coherent multiwavelength picture}
\label{sec6.4}

This work delves into the radio, optical, and X-ray properties of the S-shaped radio quasar PKS\,2300$-$18 to probe the origin and evolution of such enigmatic sources through a comprehensive multiwavelength approach. In this section, we provide a concise summary of the insights gained from our analysis of the system across various spatial scales, beginning from the vicinity of the quasar core and extending up to a few Mpc distance.

\subsubsection{Light-days to pc scale}

Optical observations revealed that the host displays a quasar spectrum characterised by double-peaked broad $H_{\alpha}$ and $H_{\beta}$ emission lines. Analysing the luminosity of the spectral lines, we were able to investigate the source on a scale of a few light days and calculate the mass of the SMBH as 2.31$\times 10^8$ M$_{\odot}$ along with the size of the BLR as 14 light-days. Given the precessing nature of the source and optical variability, we considered the idea of the double-peaked lines being due to the a binary SMBH at the centre carrying their own system of BLR clouds. However, such an analysis gave us separation between the binary SMBH's smaller or nearly the same as that of the radius of the BLRs, implying that in such a scenario both the BLRs should exist as a circumbinary BLR or as single BLR. This led us to rule out the hypothesis of a dual SMBH carrying its own BLR clouds and conclude that it is most likely the kinematics of the BLR clouds that leads to such a double-peaked spectral shape. Nevertheless, this does not dismiss the possibility of a binary SMBH system co-orbiting at the centre. Based on our calculations of the precession period, it is conceivable that a binary system exists with a separation of approximately $\sim$ $130-340$ light-days.  The variability in the spectra of the optical data and the assumptions for modelling of radio jets might be responsible for the inconsistencies in the estimated BBH separation between the optical and radio studies. Hence, to accurately determine the mechanism at the centre that results in the double-peaked broad emission-line spectrum, comprehensive studies such as spectral monitoring and reverberation mapping are essential. These methods will provide precise measurements of the central BH mass(es) and help distinguish between the presence of a dual SMBH system and the complex dynamic behaviour of BLR clouds.

Moving to the pc-scale with the milliarcsec (mas) resolution VLBI radio observations, as seen in Fig. ~\ref{fig5}, we see a single-component core of the radio source along with a one-sided jet. Using multi-epoch observations conducted over a period of 20 years, we could clearly track the positional changes of the topmost blob in five different observations and estimate the proper motion of the pc-scale radio jet as 2.3c, indicating superluminal motion. Also, from multi-epoch observation of the radio core from 5 to 40 GHz, radio variability was detected due to either some variation in the accretion rate or due to propagation of shocks in the pc-scale jets.

\subsubsection{Kpc to Mpc scale}
\label{sec6.4.2}

In the optical map shown in Fig.~\ref{fig1} (right bottom insert), the host quasar is seen merging with a companion galaxy at a separation of 14 kpc. This merger is characterised by a tidal tail and a shared envelope of dust and gas surrounding the system. Considering the possibility of the quasar already hosting an SMBH binary system at its centre, this may also make it a case of a triple merger system (\citealt{2007MNRAS.377..957H}; \citealt{2011MNRAS.416.1745D}; \citealt{liuandHo2019}); alternatively it is also possible that the host quasar captured the SMBH from the companion galaxy during a previous close interaction while merging, leading to a binary SMBH at the centre.\
In the high-resolution VLA L band map in Fig.~\ref{fig3}, we see inversion-symmetric S-shaped jets that are around 200 kpc in size displaying significant curvature/bending of jets.  The northern jet has three distinct knots, whereas the southern jet is much more diffuse, possibly due to the Dopper-boosting effect. As seen in Figs~\ref{fig2} and ~\ref{fig3}, this S-shaped jet is embedded in low surface brightness diffuse emission comprising a 5 arcmin structure that is 760 kpc in linear size, making it a giant S-shaped radio quasar. 

Given the S-shaped morphology of the source, to study the time evolution of the radio source and trace the movement of the lobes over a period of time we segregated the source into multiple circular regions, to conduct spectral ageing analysis. From the ageing analysis, the spectral age of the oldest plasma found in the diffuse wings
was $\sim$40 Myr, and the youngest plasma in the jets was $\sim$20 Myr. The
spectral ages throughout the source closely followed the gradient
seen in the spectral index map in Fig.~\ref{fig7}, with the steepest plasma found in the wings that are oriented in the northeast and southwest direction and the youngest plasma in the core and in S-shaped jets that are oriented in the north-south direction. This implies a jet reorientation taking place in the anticlockwise direction, on a timescale of a few million years (see illustration in Fig.~\ref{fig16}). The precession was also modelled using a kinematical jet precession model, where the precession was estimated to be 12$\pm$8 Myr. The discrepancy between the precession period from modelling and the oldest plasma age from the ageing analysis could hint that the source has undergone multiple cycles of precession with the oldest plasma from the oldest episode being found in the wings. It is also possible that the spectral ages represent upper limits and that the true age could be significantly less; nevertheless, this discrepancy is still considerably small. The mechanism responsible for jet precession was studied in detail, where the possibility of a binary SMBH causing geodetic precession and a tilted accretion disk remain equally plausible explanations.

The spectrum of the X-ray emission oberved by Chandra was modelled with a combination of thermal and power-law models. The hot gas detected in the central $\sim$10 kpc is likely present not only in the core, but also within the entire host galaxy. Nevertheless, in one of the models fitted to the spectrum, the metallicity of this gas was found to be as low as 3 per cent solar,
which could be a result of the photo-ionisation of the gas in the direct vicinity of the AGN/SMBHs due to quasar emission. The Chandra X-ray map in Fig.~\ref{fig14} revealed three radio knots, K1 being the most prominent. This knot is also identified on the optical map, resulting in a simultaneous multiwavelength detection of the radio knot. The X-ray emission in the knot is possibly due to the inverse Compton of CMB photons, whereas the optical emission could be due to upscattered CMB photons as a result of their interaction with the relativistic electrons present in the jet. 
In the ROSAT X-ray map, the X-ray emission extends up to a size of 2.5 Mpc. Such large-scale X-ray halos are associated with some massive galaxy clusters, e.g. the Coma cluster (\citealt{2009ApJ...696.1886B}), and can possibly be linked to cluster radio halos. However, the exact reasons for the formation of such halos remain largely unknown.

\section{Conclusion}
\label{sec7}

PKS\,2300$-$18 is a unique S-shaped radio quasar that is among the clearest examples of large-scale precessing jets in radio galaxies. The source exhibits large-scale diffuse wings along with a misaligned pair of collimated S-shaped jets, making it an excellent source to trace the morphological evolution of prcessesing jets. To analyse the source properties, we conducted multifrequency radio observations using uGMRT and the JVLA, and carried out multiwavelength investigations of the source using optical and X-ray data. We conclude that the S-shaped radio morphology is the result of continuous jet precession with a precession period of $\sim$$12\pm8$ million years, possibly caused by either a binary SMBH system at the core or a tilted accretion disc.
\begin{itemize}

\item We supplemented the flux density measurements from our dedicated observations with archival radio data, allowing us to construct spectra ranging from 183 MHz to 6 GHz. Using these spectra, we fitted a particle injection model to different sections of the wings and the jet to determine their break frequency values.

\item In the polarisation studies, it was observed that the entire source was strongly polarised, with the central part showing a high degree of polarisation. Additionally, the fractional polarisation was found to be higher in the wings.

\item In the X-ray studies using Chandra, the host galaxy and its hot interstellar medium were detected. The core spectrum was modelled with both a thermal and a power-law component. Furthermore, a strong X-ray knot coinciding with the radio and optical knot K1 was discovered, along with two fainter X-ray knots corresponding to the radio knots K2 and K3.

\item Variability was detected in the radio core observations over multiple epochs. The observed variability, which occurred over a span of a few days to weeks, was attributed to changes in the accretion rate or changes in the pc-scale jets close to the radio core.

\item The broad, double-peaked optical emission lines from the quasar spectrum were fitted with Lorentzian profiles, which yielded a central SMBH mass of $2.31\times 10^8$ M$_{\odot}$ and a BLR size of 14 light-days. This double-peaked emission line spectrum indicated complex gas kinematics within the quasar's BLR.

\item The jets and the diffuse emission of the source were modelled using a kinematical jet precession model, which determined the precession period to be $12\pm8$ million years.

\item The spectral age of the oldest plasma found in the diffuse wings was $\sim$40 Myr and that of the youngest plasma in the jets $\sim$20 Myr. The spectral ages throughout the source closely followed the gradient observed in the spectral index map, with the steepest plasma located in the wings oriented along northeast and southwest directions, and the youngest plasma found in the core and the S-shaped jets oriented along north-south direction. 

\item The source morphology results from jet reorientation taking place in the anti-clockwise direction over a span of a few million years. The discrepancy between the precession period and the maximum plasma age suggests that the source may have undergone multiple precession cycles.\\

The multiwavelength study of the S-shaped radio quasar PKS\,2300$-$18 presented in this work is among the most extensive investigations on such a select class of radio galaxies. Studying these sources is vital for an in-depth understanding of galaxy evolution through nuclear activity, as changes at the scale of the central AGN manifest as large-scale perturbations, seen in the case of radio jets. This can provide novel insights into the dynamic behaviour of the central SMBH. Multiwavelength studies also play a key role in building a comprehensive picture of the host galaxy and the central SMBH, from subpc to Mpc scales. Given that these objects are also prime candidates for hosting BBH at their centre, in their final evolutionary stages, they are expected to generate nanohertz gravitational wave (GW) emissions. Such emissions can be observed by pulsar timing arrays, which regularly monitor millisecond pulsars \citet{2023ApJ...952L..37A} to track GWs from SMBH mergers, and also by other novel methods, as suggested by \citet{2024NatAs.tmp..160S}. Consequently, radio galaxies exhibiting S-shaped morphology can serve as an ideal resource for identifying BBHs, investigating galaxy mergers, and analysing dynamic behaviour of AGNs.

\end{itemize}

\section{Acknowledgements}

The authors thank the anonymous reviewer for her/his valuable comments and Urszula Pajdosz-Śmierciak for her contribution to the radio proposals. DKW would like to thank Agnieszka Ku\'{z}micz and Bo\.{z}ena Czerny for discussions. AM would like to thank Rubinur Khatun, Martin Hardcastle, Kamil Wolnik and Subhrata Dey for scientific discusions. MJ acknowledges access to the SYNAGE software kindly provided by Matteo Murgia.  We acknowledge support from the National Science Centre research grants 2021/43/B/ST9/03246 [DKW, AM]. 

The GMRT is a national facility operated by the NCRA, TIFR. The National Radio Astronomy Observatory is a facility of the National Science Foundation operated under a cooperative agreement by Associated Universities Inc. The National Radio Astronomy Observatory running the VLA is a facility of the National Science Foundation operated under a cooperative agreement by Associated Universities, Inc. The Astrogeo VLBI FITS image database is maintained by L. Petrov, and the particular maps were contributed by Y. Y. Kovalev, A. Pushkarev, and L. Petrov.

The Pan-STARRS1 Surveys (PS1) and the PS1 public science archive have been made possible through contributions by the Institute for Astronomy, the University of Hawaii, the Pan-STARRS Project Office, the Max-Planck Society and its participating institutes, the Max Planck Institute for Astronomy, Heidelberg and the Max Planck Institute for Extraterrestrial Physics, Garching, The Johns Hopkins University, Durham University, the University of Edinburgh, the Queen's University Belfast, the Harvard-Smithsonian Center for Astrophysics, the Las Cumbres Observatory Global Telescope Network Incorporated, the National Central University of Taiwan, the Space Telescope Science Institute, the National Aeronautics and Space Administration under Grant No. NNX08AR22G issued through the Planetary Science Division of the NASA Science Mission Directorate, the National Science Foundation Grant No. AST-1238877, the University of Maryland, Eotvos Lorand University (ELTE), the Los Alamos National Laboratory, and the Gordon and Betty Moore Foundation.

This research has made use of data obtained from the Chandra Data Archive and the Chandra Source Catalog, and software provided by the Chandra X-ray Center (CXC) in the application packages CIAO and Sherpa. This research has made use of the NASA/IPAC Extra-
galactic Database (NED), which is funded by the Na-
tional Aeronautics and Space Administration and oper-
ated by the California Institute of Technology.

ZTF is a fully-automated, wide-field survey aimed at a systematic exploration of the optical transient sky. Supported by the National Science Foundation under Grants No. AST-1440341 and AST-2034437 and a collaboration including current partners Caltech, IPAC, the Oskar Klein Center at Stockholm University, the University of Maryland, University of California, Berkeley , the University of Wisconsin at Milwaukee, University of Warwick, Ruhr University, Cornell University, Northwestern University and Drexel University. Operations are conducted by COO, IPAC, and UW.

\section*{Data Availability}

The data underlying this paper will be shared on reasonable request to the corresponding author.



\bibliographystyle{mnras}
\bibliography{reference} 

\label{lastpage}

\end{document}